\documentclass[traditabstract]{aa}  
\usepackage{graphicx,txfonts,natbib,amssymb,psfig}

\def\deg2{{\rm~deg}^2}
\def\degm2{\rm~deg^{-2}}
\def\arc2{{\rm~arcmin}^2}
\begin{document}
   \title{The WIRCam Deep Survey I: Counts, colours and mass-functions derived from near-infrared imaging in the CFHTLS deep fields \thanks{Based on observations obtained with
    MegaPrime/MegaCam, a joint project of CFHT and CEA/DAPNIA, at the
    Canada-France-Hawaii Telescope (CFHT) which is operated by the
    National Research Council (NRC) of Canada, the Institut National
    des Sciences de l'Univers of the Centre National de la Recherche
    Scientifique (CNRS) of France, and the University of Hawaii. This
    work is based in part on data products produced at TERAPIX and the
    Canadian Astronomy Data Centre as part of the Canada-France-Hawaii
    Telescope Legacy Survey, a collaborative project of NRC and
    CNRS.} } \offprints {R.\ Bielby}   \date{Received date / 
  Accepted date}\titlerunning{WIRDS I: Counts, colours and mass functions}
\authorrunning{Bielby et al.} 

   \author{R. Bielby\inst{1,2}, P. Hudelot\inst{1}, H. J. McCracken\inst{1}, O. Ilbert\inst{3}, E. Daddi\inst{4}, O. Le F{\`e}vre\inst{3},\\ V. Gonzalez-Perez\inst{2}, J.-P. Kneib\inst{3}, C. Marmo\inst{1,5}, Y. Mellier\inst{1}, M. Salvato\inst{6}, D. B. Sanders\inst{7}, C. J. Willott\inst{8} \fnmsep}

   \institute{Institut dAstrophysique de Paris, UMR7095 CNRS, Universit\'e  Pierre et Marie Curie, 98 bis Boulevard Arago, 75014 Paris, France.
       	\and
       Dept. of Physics, Durham University, South Road, Durham, DH1 3LE, UK.
       \and
       Laboratoire d'Astrophysique de Marseille, Universit\'e Aix-Marseille, 38 rue Fr\'ed\'eric Joliot-Curie, 13388 Marseille, France.
       \and
       Service d'Astrophysique, CEA/Saclay, 91191 Gif-sur-Yvette, France.
       \and
	   Laboratoire IDES, UMR 8148 CNRS, UniversitŽ Paris-Sud XI, 91405 Orsay, France.
       \and
       Max Planck Institut f\"ur Plasma Physik and Excellence Cluster, Boltzmannstrasse 2, 85748 Garching, Germany.
       \and
       Institute for Astronomy, 2680 Woodlawn Dr., University of Hawaii, Honolulu, HI 96822, USA.
       \and	
       Herzberg Institute of Astrophysics, National Research Council, 5071 West Saanich Road, Victoria, BC V9E 2E7, Canada.\\
       \email{rmbielby@gmail.com} } \abstract{We present a new near-infrared imaging survey in the four CFHTLS deep fields: the WIRCam Deep Survey or ``WIRDS''. WIRDS comprises extremely deep, high quality (FWHM $\sim0.6\arcsec$) $J$, $H$ and $K_{\rm s}$ imaging covering a total effective area of $2.1\deg2$ and reaching AB 50\% completeness limits of $\approx24.5$. We combine our images with the CFHTLS to create a unique eight-band $ugrizJHK_S$ photometric catalogues in the four CFHTLS deep fields; these four separate fields allow us to make a robust estimate of the effect of cosmic variance for all our measurements. We use these catalogues in combination with $\approx9,800$ spectroscopic redshifts to estimate precise photometric redshifts ($\sigma_{\Delta z/(1+z)}\lesssim0.03$ at $i<25$), galaxy types, star-formation rates and stellar masses for a unique sample of $\approx1.8$ million galaxies. Our $JHK_{\rm s}$ number counts are consistent with previous studies. We apply the ``$BzK$'' selection to our $gzK$ filter set and find that the star forming $BzK$ selection successfully selects 76\% of star-forming galaxies in the redshift range $1.4<z<2.5$ in our photometric catalogue, based on our photometric redshift measurement. Similarly the passive $BzK$ selection returns 52\% of the passive $1.4<z<2.5$ population identified in the photometric catalogue. We present the mass functions of the total galaxy population as a function of redshift up to $z=2$ {and present fits using double Schechter functions. A mass-dependent evolution of the mass function is seen with the numbers of galaxies with masses of $M\lesssim10^{10.75}$ still evolving at $z\lesssim1$, but galaxies of higher mass reaching their present day numbers by $z\sim0.8-1$. This is consistent with the present picture of downsizing in galaxy evolution. We compare our results with the predictions of the GALFORM semi-analytical galaxy formation model and find that the simulations provide a relatively successful fit to the observed mass functions at intermediate masses (i.e. $10\lesssim{\rm log}(M/M_\odot)\lesssim11$). However, as is common with semi-analytical predictions of the mass function, the GALFORM results under-predict the mass function at low masses (i.e. ${\rm log}(M/M_\odot)\lesssim10$), whilst the fit as a whole degrades beyond redshifts of $z\sim1.2$.} All photometric catalogues and images are made publicly available from TERAPIX and CADC.}

   \keywords{cosmology: observations --- cosmology: dark matter --- galaxies:
   large scale structure of the universe --- galaxies: surveys}
\authorrunning{R. Bielby et al}
   \maketitle
%

\section{Introduction}

\setcounter{footnote}{0}

Today, observational evidence concurs that star formation in the Universe reached its peak within the redshift range $1<z<2$, whilst $\sim 50\%-70\%$ of mass assembly took place in the redshift range $1<z<3$ \citep{1997ApJ...486L..11C,2003ApJ...587...25D,2007A&A...476..137A, 2007A&A...474..443P, 2007ApJ...660L..43N}. In addition, observations have revealed the existence of a substantial population of very massive galaxies at $z>1$ \citep[e.g.][]{mullis05,stanford05,stanford06,2010ApJ...718...23S,2011NJPh...13l5014F}. Such observations have acted to stimulate significant progress in our understanding of galaxy formation processes, in particular highlighting the importance of feedback processes in shaping the pace and form of the growth of galaxies. Semi-analytical simulations have shown that the addition of AGN feedback allow such models to predict both the substantial population of massive galaxies at $z>1$ and the rise in the star-formation density to $z\sim2$, whilst simultaneously matching the luminosity function and colour distribution of galaxies in the local Universe \citep[e.g.][]{2006MNRAS.370..645B,2006MNRAS.365...11C,2006MNRAS.366..499D,2007MNRAS.375....2D,2008MNRAS.384....2G,2011MNRAS.413..101G}.

Although models (and hence our understanding of galaxy evolution) are continually improving and have had many successes, there are still limitations and inconsistencies. For example, semi-analytic models have difficulty simultaneously producing the total mass function of the massive galaxy population at $z>1$ and stellar growth in highly star-forming submillimetre galaxies observed at $z\sim1$ \citep[e.g.][]{2005MNRAS.356.1191B,2006MNRAS.370..645B}. In addition, there remain instances of observations challenging the model predictions \citep[e.g.][]{2010ApJ...708..202M,2011MNRAS.413..101G,2012MNRAS.421.2904H}, which highlight that our understanding of galaxy formation remains incomplete. Given the central role that intermediate redshift populations play in our understanding of galaxy formation and evolution as a whole, gathering deeper wide-area observations at this redshift range is of crucial importance.

Below $z\sim1$, the galaxy population can be relatively easily identified via the optical features such as the 4000\AA\ break and a variety of absorption and emission lines. Above $z\sim2$, the ultraviolet spectral features are redshifted into the optical and the galaxy population can be readily identified via the Lyman Break \citep[e.g.][]{steidel96,steidel03,2006ApJ...652..994C,2011MNRAS.414....2B} and Ly$\alpha$ emission \citep[e.g.][]{1998AJ....115.1319C,2003ApJ...582...60O,2006ApJ...642L..13G}. However, to follow the galaxy population into the redshift range $1\lesssim z\lesssim2.5$, near-infrared observations are essential as all spectral features move out of visible bands. Furthermore, the role of environment and galaxy formation process at these redshifts is largely unexplored \citep{2009arXiv0906.4662R}. Near-infrared galaxy samples offer several well-known advantages compared to purely optical selections (see, for example \cite{1994ApJ...434..114C}); at $z\sim1$ near-infrared selected galaxies are seen in the rest-frame optical. This corresponds more closely to a stellar-mass-selected sample and are therefore less prone to the uncertain effects of dust extinction present in these redshifts for rest-frame optically selected samples. As the $K_{\rm s}$ band $k-$ corrections are insensitive to galaxy type over a wide redshift range, these samples provide a fairly unbiased census of galaxy populations at high redshifts (providing that the extinction is not too high, as in the case of some submillimeter galaxies). Such samples represent the ideal input catalogues from which to extract targets for spectroscopic surveys as well as for determining accurate photometric redshifts and making comparisons with models. 

\cite{1996AJ....112..839C} carried out one of the first extremely deep, complete $K$ selected spectroscopic surveys and showed that star-forming galaxies at low redshifts have smaller masses than actively star-forming galaxies at $z\sim1$, a phenomenon known as ``downsizing''. Stated another way, the sites of star-formation ``migrate'' from higher-mass systems at high redshift to lower-mass systems at lower redshifts; this anti-hierarchical behaviour seemed at odds with the hierarchical picture of galaxy formation and helped to stimulate refinement of theoretical models of galaxy formation to account for these observations. More recently, the K20 survey \citep{2002A&A...381L..68C} reaching $K_{\rm s}\approx 21.8$ and the GDDS survey \citep{2004AJ....127.2455A} reaching $K\approx 22.4$ provided further observations to test the existing knowledge and understanding of the galaxy formation process. The areas covered by these surveys was small, comprising only $\sim 55$ arcmin$^2$ and $\sim 30$ arcmin$^2$ in K20 and GDDS respectively. Improvements in recent years have been provided by the Ultra Deep Survey (UDS) segment of the United Kingdom Infrared Telescope (UKIRT) Infrared Deep Sky Survey \citep[UKIDSS;][]{lawrence07,2007MNRAS.379L..25L}, which covers an area of 0.77 deg$^2$ to a depth of $K\sim25$ (AB) and the Cosmic Evolution Survey \citep[COSMOS;][]{2007ApJS..172....1S,koekemoer07} which covers a total area of 2 deg$^2$ to a depth of $K=24$ (AB). Each of these surveys covers just a single field however, and cosmic variance effects are particularly important, especially for the red galaxy populations that are dominated by highly-clustered objects.

A number of studies have analysed the evolution of the build-up of stellar mass in galaxies via the stellar mass function, finding that the stellar mass density approximately doubles between $z\sim1$ and $z\sim0$ \citep[e.g.][]{2005ApJ...625..621B,2006ApJ...651..120B,2006A&A...453..869B,2006ApJ...639L...1P}. \citet{2010ApJ...709..644I} performed similar analyses out to redshifts of $z\sim2$ and found that, for quiescent galaxies, $z\sim1$ represents an epoch of transition in their stellar mass assembly, showing that the stellar mass density for the quiescent population increases by over an order of magnitude from $z\sim2$ to $z\sim0.8$, but only increases by a factor of 2 between $z\sim0.8$ and the present day. Combining this result with the morphologies of the $z\lesssim0.8$ quiescent sample, they concluded that a dominant mechanism links the shutdown of star formation and the acquisition of an elliptical morphology in massive galaxies. Although these studies have provided significant insights into the evolution of the galaxy population, building on these results requires further wide field deep NIR imaging to probe a fuller range of the stellar mass content across cosmic time. 

In this paper, the first of a series, we present such a survey: the WIRCam Deep Survey or ``WIRDS''. We describe the field layout, observing strategy, data processing techniques and present a comprehensive quality assessment of the final released imaging data. These data have already been used in a number of studies: \citealt{2010A&A...523A..66B,2011A&A...525A.143C,2011A&A...529L...5A,2012A&A...539A..31C,2012A&A...537A..88R,2011ApJ...735...86W,2012arXiv1202.5330W}. Both images and catalogues are publicly available from CADC\footnote{{http://cadcwww.dao.nrc.ca/cfht/WIRDST0002.html}} and TERAPIX\footnote{http://terapix.iap.fr/article.php?id\_article$=$832)} data centers. In our second paper, we will describe clustering measurements for mass-selected samples derived using these catalogues. 

This paper is organised as follows: Section~\ref{sec:obs-data} presents the survey strategy, observations, data reductions and quality assessments; Section~\ref{sec:bzk} we present galaxy number counts and sub-classes of objects selected using various selection criteria in joint CFHTLS-WIRDS dataset, including $BzK$ selections and separation by type based on rest-frame UV colours. In Section~\ref{sec:massfunc} we present an analysis of the mass completeness of the data and present precise mass functions measured to a redshift of $z=2$. Section~\ref{sec:conclusions} provides a summary and our conclusions. The WIRDS survey is designed to reach AB magnitudes of 24 in all near-infrared bands. The science rationale for this was the following: firstly, for the purposes of deriving accurate photometric redshifts below the typical spectroscopic limit of $i=24.5$; and secondly, to reach at least to $M^\star$ for galaxies up to $z=3$.

Throughout this paper, all magnitudes are given in the AB system unless stated otherwise. Unless otherwise stated, we use a $\Lambda$CDM cosmology given by $H_0=70~\mbox{km/s/Mpc}$, $\Omega_0=0.3$ and $\Lambda_0=0.7$.

\section{Observations and data reductions}
\label{sec:obs-data}

\subsection{Observations}

The NIR observations were taken using the WIRCam detector \citep{puget04} at the Canada France Hawaii Telescope (CFHT), with the exception of D2 $J$ which was observed with WFCAM on UKIRT. WIRCam consists of four $2048\times2048$ HgCdTe arrays arranged in a $2\times2$ format, with gaps of $45\arcsec$ between adjacent chips. The detector pixel scale is $0.3\arcsec$/pixel resulting in a field-of-view of $21\arcmin\times21\arcmin$. Observations were taken in a series of runs from 2005-2007 and were made in co-ordination with the COSMOS consortium. In this work we combine the NIR imaging with the CFHTLS T0006 imaging described by \citet{CFHTLS_T0006}. We note that in the D2 field, we include the COSMOS data that has already been described by \citet{2010ApJ...708..202M} although we restrict ourselves only to the central $1\deg2$ and use this data in conjunction with the CFHTLS optical $ugriz$ data (rather than the COSMOS Subaru observations) in order to be consistent between each of the four fields.  

\begin{table*}
\caption{Overview of the WIRDS/COSMOS observations.}             
\label{table:observations}      
\centering          
\begin{tabular}{l c c c c c c c c}
\hline\hline       
Field & Subfields & R.A. & Dec & Area (masked) & Band & Ind. exp. &$N_{im}$ & Mean exp. time\\ 
      &           & \multicolumn{2}{c}{(J2000)}  &($\deg2$)&  & (s)  &       & (s) \\ 
\hline                    
CFHTLS D1  & 5 & 02:26:00 &-04:30:00& 0.6 (0.49)&$J$& 45&1,460 & 13,140 \\  
       &  & && 0.6 (0.49)&$H$&15 & 4,916& 14,748\\  
       &  & && 0.6 (0.49)&$K_{\rm s}$&20 & 4,235& 16,940\\  
\hline                    
CFHTLS D2/COSMOS &9& 10:00:29 &+02:12:21& 1.0 (0.80)&$J$& 5/10 & 15,604/5,447 & 13,710 \\  
      & &&& 1.0 (0.80)&$H$& 15 & 14,000 & 13,125\\ 
      && & &1.0 (0.80)&$K_{\rm s}$& 20 & 7,272 & 9,090 \\ 
\hline                  
CFHTLS D3/EGS &3& 14:17:54 & +52:30:31 & 0.4 (0.39)&$J$&45 & 1,079 & 16,185\\  
      && &&0.4 (0.39)&$H$& 15& 3,390 &16,950\\  
      && &&0.4 (0.39)&$K_{\rm s}$& 20& 2,622 &17,480\\  
\hline                    
CFHTLS D4/Q2215-1744  &3& 22:15:31 & -17:44:05.4 & 0.4 (0.36)&$J$&45 & 1,258 & 18,870\\
      && &&0.4 (0.36)&$H$& 15& 3,166 & 15,830\\  
      && &&0.4 (0.36)&$K_{\rm s}$& 20& 2,362&15,746\\  
\hline                  
\end{tabular}
\end{table*}

Observation are summarised in Table~\ref{table:observations}. The mean exposure time listed is approximately the exposure time per pixel and is simply the individual exposure time multiplied by the number of exposures and divided by the number of subfields across the field. Our observing strategy was chosen in order to ensure uniform coverage reaching a target depth of 24AB. Given the limited observing time allocated to the program, this precluded full coverage of each of the  $1^\circ\times1^\circ$ CFHTLS deep fields. Instead a subsection of each of the D1, D3 and D4 fields was chosen (however D2-COSMOS was covered in its entirety by the COSMOS consortium). Five WIRCam pointings were made in the D1 field and three in D3 and D4 fields. This gives a total area of $0.6\deg2+1.0\deg2+0.4\deg2+0.4\deg2=2.4\deg2$ (or $2.03\deg2$ after masking). The subfields in D1 field were chosen to overlap with the VIMOS-VLT Deep Survey \citep[VVDS; ][]{2005A&A...439..845L} and the Spitzer Wide-area Infrared Extragalactic Legacy Survey \citep[SWIRE; ][]{2003PASP..115..897L,2004ApJS..154...54L}, whilst the entire field overlaps with a portion of the XMM Large Scale Structure Survey \citep[XMM-LSS; ][]{2004JCAP...09..011P}. Two of the pointings in the D3 field were chosen to provide near-infrared imaging for a segment of the DEEP2 spectroscopic survey and the AEGIS X-ray survey. Finally, the three D4 pointings were chosen to coincide with a concurrent program to obtain spectroscopic redshifts of $z\approx4$ galaxies using VLT VIMOS (PI: J. Bergeron, 077.A-0357).

Observations were conducted using $J$, $H$ and $K_{\rm s}$ filters. Transmission plots of these filters are available from CFHT\footnote{cfht.hawaii.edu/Instruments/Filters/wircam.html}. The integration times per exposure for all $J$, $H$ and $K_{\rm s}$ band observations was 45s, 15s and 20s respectively.
The observations were carried out in queue-scheduled mode with image quality constraints of $0.55\arcsec<FWHM<0.65\arcsec$ and were ``micro-dithered'' (the telescope was displaced using sub-pixel offsets) using the standard WIRCam micro-dither pattern consisting of a 2$\times$2 dither patter with offsets between consecutive dithers of 0.5 pixels. Since WIRCam has a pixel scale of $0.3\arcsec$/pixel, this micro-dithering is required in order to produce well-sampled images under our seeing constraint and also to allow matching with the CFHTLS pixel-scale of $0.186\arcsec$/pixel. A further large-scale ($\approx1-2\arcmin$) dithering pattern was applied to the observations to avoid gaps in the coverage due to the gaps between adjacent CCDs.


\subsection{Data reduction}
\label{sec:datared}

\subsubsection{WIRDS D1, D3 and D4 fields}

The WIRDS data set were reduced at CFHT and TERAPIX. Initial processing was made at CFHT using the \texttt{I'iwi}\footnote{cfht.hawaii.edu/Instruments/Imaging/WIRCam/IiwiVersion1Doc.html} preprocessing pipeline, which incorporates detrending and initial sky-subtraction processing. In order to ensure precise measurements of fluxes in background-dominated near-infrared observations like WIRDS, we use a two-step reduction process in which objects were flagged in initial stacks created using a standard sky-subtraction procedure. These objects were identified and masked in a second-pass sky-subtraction. The overall reduction process was as follows:

\begin{enumerate}
\item Bias subtraction and flat-fielding (detrending)
\item Initial sky-subtraction using object mask generation from the individual images
\item Cross-talk correction
\item Astrometric calibration and production of initial pre-processed stacks
\item Second-pass sky-subtraction using object mask from the pre-processed stack and individual images
\item Cross-talk correction
\item Astrometric and photometric calibration
\item Production of final processed image stacks
\end{enumerate}

These steps are now described in detail. The detrending stage involves the treatment of a number of instrumental imprints: flagging saturated pixels, non-linearity correction, reference pixel subtraction, dark-frame subtraction, dome flat-fielding, bad-pixel masking and guide window masking. The pre-processing of the images consists of bias-subtraction and flat-fielding using the bias and flat-field frames provided by the CFHT WIRCam queue observing team. A global bad pixel mask was generated using the flat to identify the dead pixels and the dark frames to identify hot pixels. For each image, we used the TERAPIX tool \texttt{QualityFITS} to produce weight-maps, object catalogues and overviews of individual image qualities (e.g. seeing, depth). The production of the weight maps made use of the \texttt{WeightWatcher} software \citep{weightwatcher}.

After detrending, the initial sky-subtraction process was made. Median skies were produced for each individual image using up to 30 (object masked) images (equivalent to $\sim10$mins total exposure time) in a given dithering pattern with on-sky separations of $<30\arcsec$. The images were then sky-subtracted using these sky frames. Sky-flat images were constructed for each image using combinations of masked adjacent images taken over time intervals $\Delta t$ and angular separations of  $\Delta \theta$ of $\Delta t<20~\rm{mins}$ and $\Delta\theta<10\arcmin$ respectively. In general individual sky-flat images consisted of $\approx8-9$ multi-extension fits image files. To perform the masking of the images prior to constructing the sky-flats, \texttt{SExtractor} \citep{sextractor} was used to identify objects using a detection threshold of $1.5\sigma$, a minimum detection area of four pixels, a background mesh size of 64 pixels and a background filter size of one pixel. A ``check-image'' of type ``\texttt{OBJECTS}'' was generated to identify objects in each image. With the images masked, the sky-flats were constructed and subtracted from the original images.

After sky-subtraction, we cross-talk corrected our images. Cross-talk arises as a repetitive signal across the WIRCam amplifiers on a given array and although cross-talk signals have been dealt with at the hardware level on WIRCam since 2008, many of our observations were taken prior to this date and so it must be dealt with in software. The cross-talk appears as ``ghosts'' of bright objects in adjacent amplifiers in a given array in the WIRCam detector. Depending on the form, the effect can be seen in all 32 amplifiers in a given array (where each amplifier consists of $64\times2048$ pixels). We attempt to correct for the effect using a form of the ``Medamp'' procedure, where object masks were first created for each image using \texttt{SExtractor}. For each array, we then take the median of the 32 amplifiers as a way of identifying those features that are common to all 32 amplifiers (i.e. the cross-talk signals). These signals are then subtracted from the original image.

Astrometric calibration was made using \texttt{Scamp} \citep{scamp} with 2MASS as the astrometric reference catalogue. We obtained astrometric solutions with internal accuracies of $\approx0.1\arcsec$. Initial median-combined stacks were constructed using \texttt{Swarp} \citep{swarp} at the instrument pixel scale. From the initial stacks we constructed a full object mask incorporating the fainter objects not detectable in individual images using the same time techniques for the individual object masks. 
Following this, the sky flat creation was repeated on the original detrended images, using both the individual image masks and the stack-mask to mask objects from the images before making sky-flats. The new sky flats were then subtracted from the detrended images. Cross-talk correction was again performed on the images. Astrometric calibration was performed using \texttt{Scamp} and the photometric calibration was then performed by matching to 2MASS photometry. All images were scaled to have a zero-point of 30.00AB.

\begin{figure}[!h]
\centering
\includegraphics[width=90.mm]{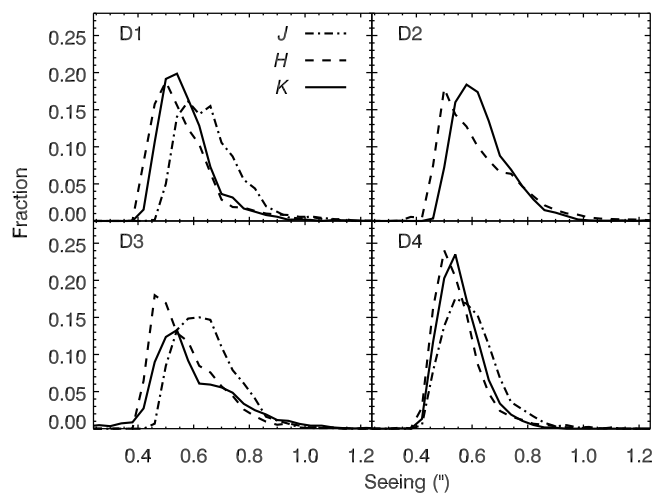}
   \caption{Flux\_radius distribution of the individual exposures in each of the four fields. The dashed-dotted, dashed and solid lines show $JHK_{\rm s}$-bands respectively (note that the COSMOS $J$-band data is not shown here, as we do not have access to the individual images used in the stack).}
   \label{fig:seeing_hist}
\end{figure}

Before the final stack, we performed quality assessment of the individual images. First each image was inspected and those with severe defects were removed. We measured the seeing for each image based on the flux radius using \texttt{PSFEx} (this is defined as the radius which includes 50\% of the flux with respect to the total flux) . All the final stacked images were combined after discarding the 5\% worst-seeing in each field and filter combination. The distributions of the flux\_radius measurements for all images for each field/filter combination are shown in Fig.~\ref{fig:seeing_hist}. In each case, the dash-dotted, dashed and solid lines line shows the flux\_radius distribution across all $JHK_{\rm s}$ respectively (as a fraction of the total number of images). Distributions are sharply peaked with a small tail of poor-seeing images (which are removed from our final sample by our 5\% cut). The 5\% cuts for each field/filter combination and the measured seeing for each final stacked image are given in Table~\ref{tab:psf}.

\begin{table}[!h]
\caption{Flux-radius properties of the WIRDS data. The seeing cuts give the upper limit above which individual exposures were rejected prior to stacking the images using \texttt{Swarp}.}             
\label{tab:psf} 
\centering          
\begin{tabular}{l c c c c c c}     
\hline   
Field & \multicolumn{3}{c}{Seeing cut} & \multicolumn{3}{c}{Final flux\_radius} \\ 
          & $J$ & $H$ & $K_{\rm s}$ & $J$ & $H$ & $K_{\rm s}$    \\      
\hline                    
D1        & 0.90\arcsec& 0.87\arcsec & 0.91\arcsec& 0.68\arcsec& 0.62\arcsec& 0.67\arcsec  \\      
D2        &  & 1.1\arcsec & 1.1\arcsec &0.91\arcsec & 0.69\arcsec& 0.68\arcsec  \\      
D3        & 0.82\arcsec & 0.81\arcsec & 0.89\arcsec &0.64\arcsec & 0.58\arcsec& 0.60\arcsec  \\      
D4        & 0.77\arcsec & 0.72\arcsec & 0.70\arcsec &0.59\arcsec & 0.57\arcsec& 0.58\arcsec  \\      
\hline                  
\end{tabular}
\end{table}

Before stacking, images were resampled using a Lanczos-2 4-tap filter with a 128 pixels mesh for background subtraction. Stacks were created in each band by sigma-combining appropriately weighted pixels using a modified version of \texttt{Swarp} kindly supplied by S. Foucaud; the pixel rejection limit was set to $3\sigma$. Images were rescaled to match the pixel scale of $0.186\arcsec/$pixel and tangent points of the CFHTLS deep stacks, making it easy to extract matched catalogues in double image mode. We note that at this point we performed a qualitative check on the astrometry of the images by constructing combined colour images of the final stacks in each fields using \texttt{Stiff} package \citep{stiff}.

\subsubsection{D2/COSMOS data reductions}

The COSMOS $H$ band data was reduced identically as the other fields described above. However, the WIRCam $K_{\rm s}$ and WFCAM $J$ band data were processed using a different pipeline which is described in \citet{2010ApJ...708..202M}. The WFCAM $J$ band data is partially described in \citet{2011ApJ...730...68C}.

\subsection{Completeness and coverage}

\begin{figure*}
\centering
\includegraphics[width=0.32\textwidth]{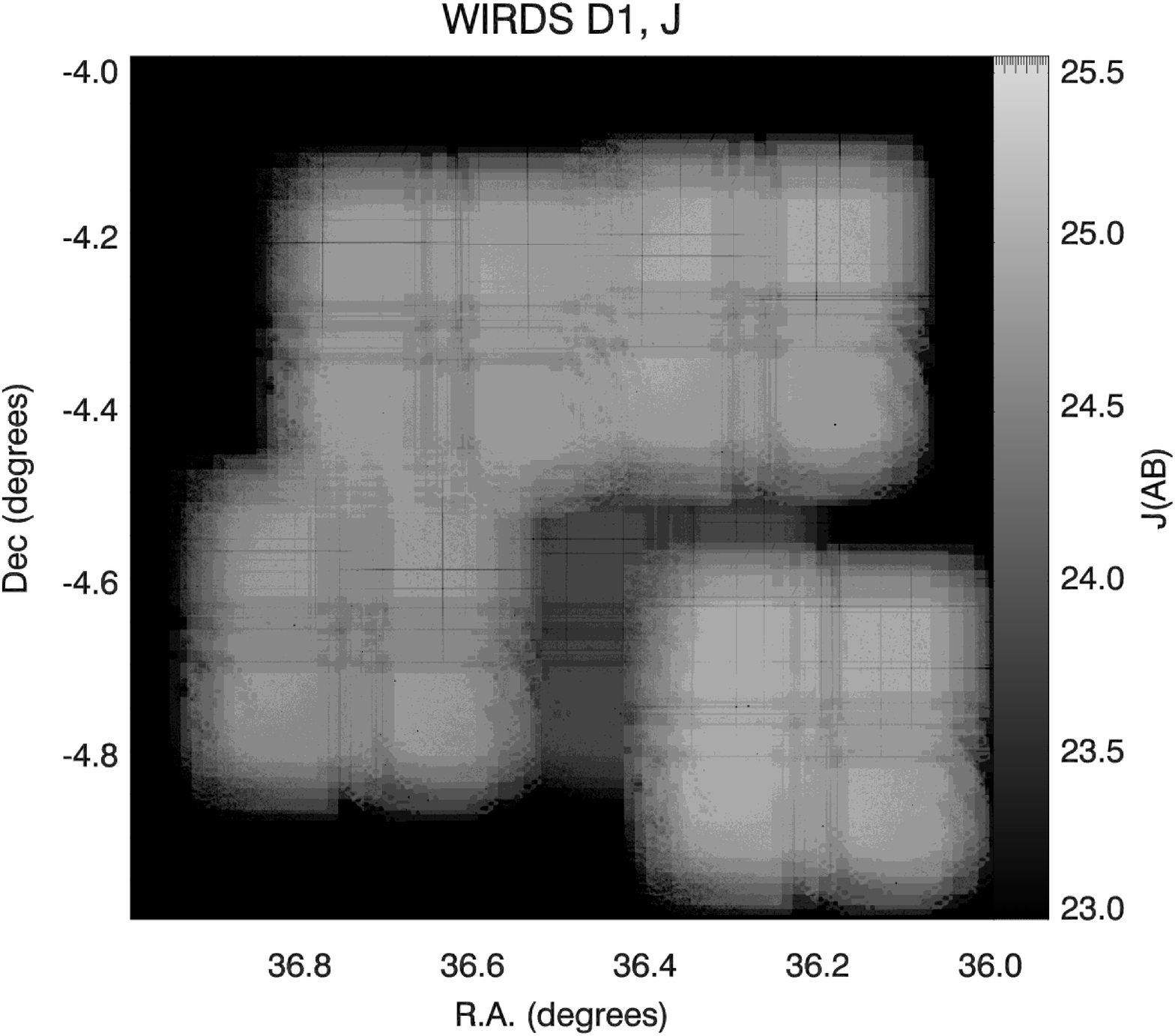}
\includegraphics[width=0.32\textwidth]{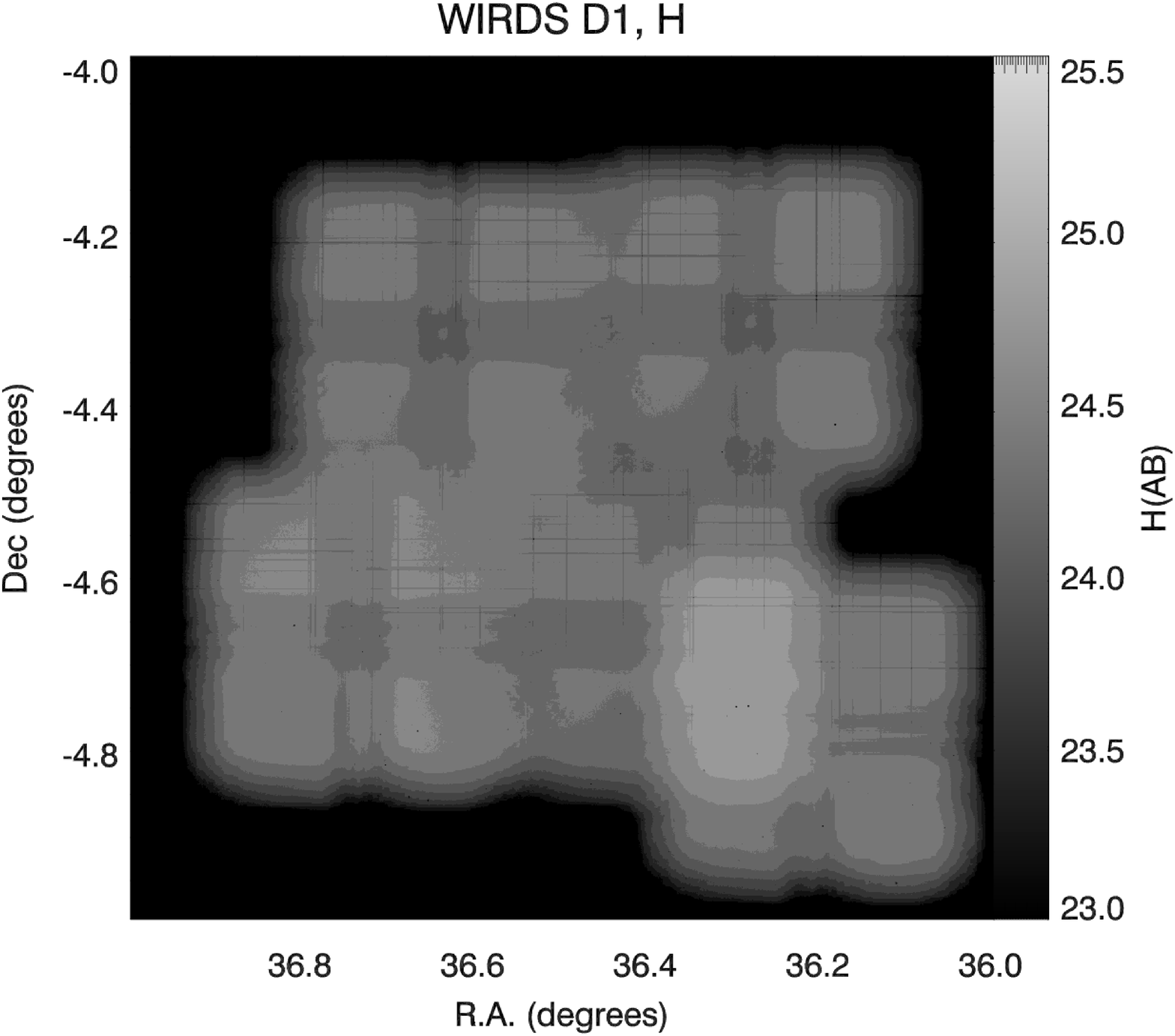}
\includegraphics[width=0.32\textwidth]{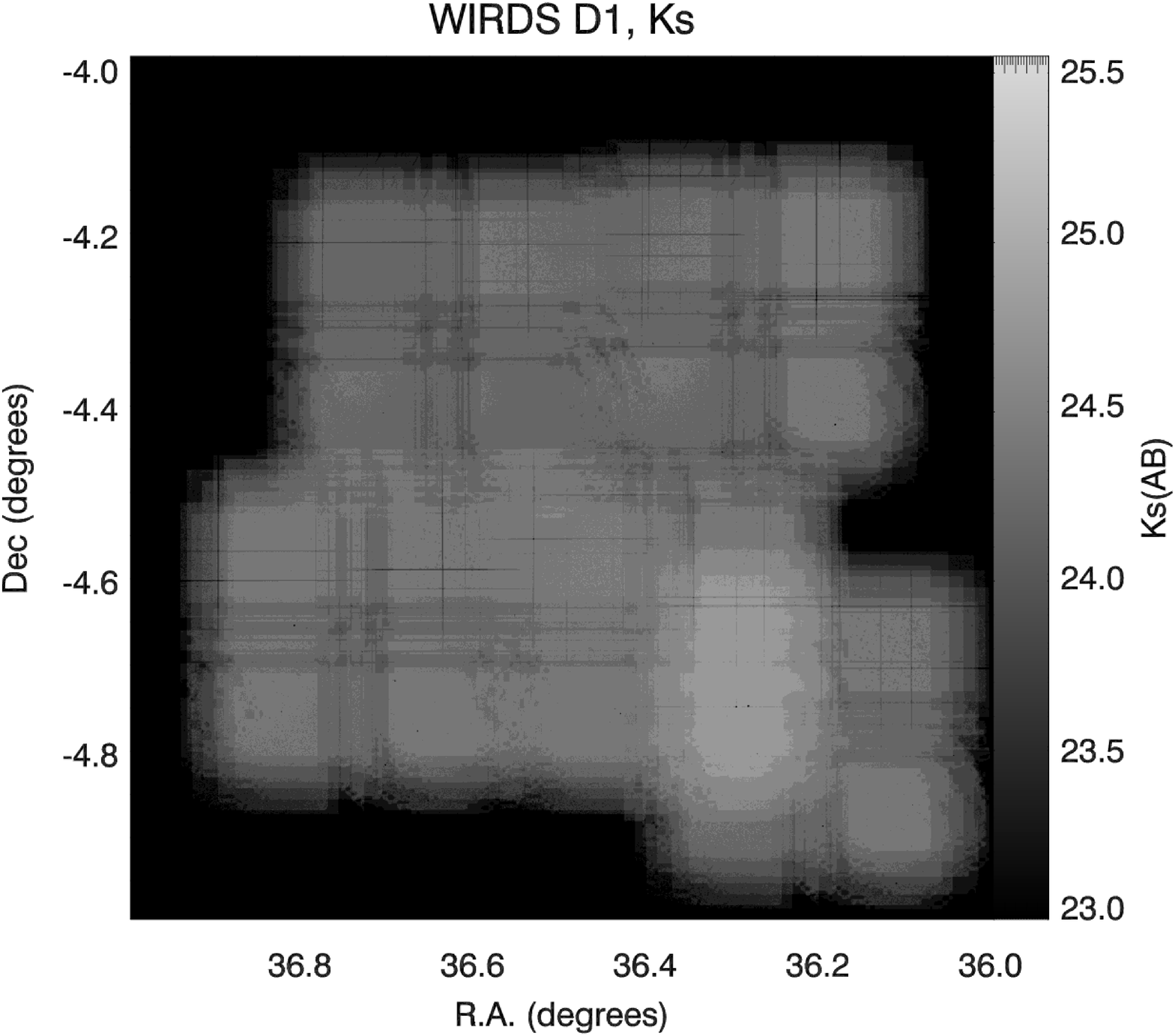}

\includegraphics[width=0.32\textwidth]{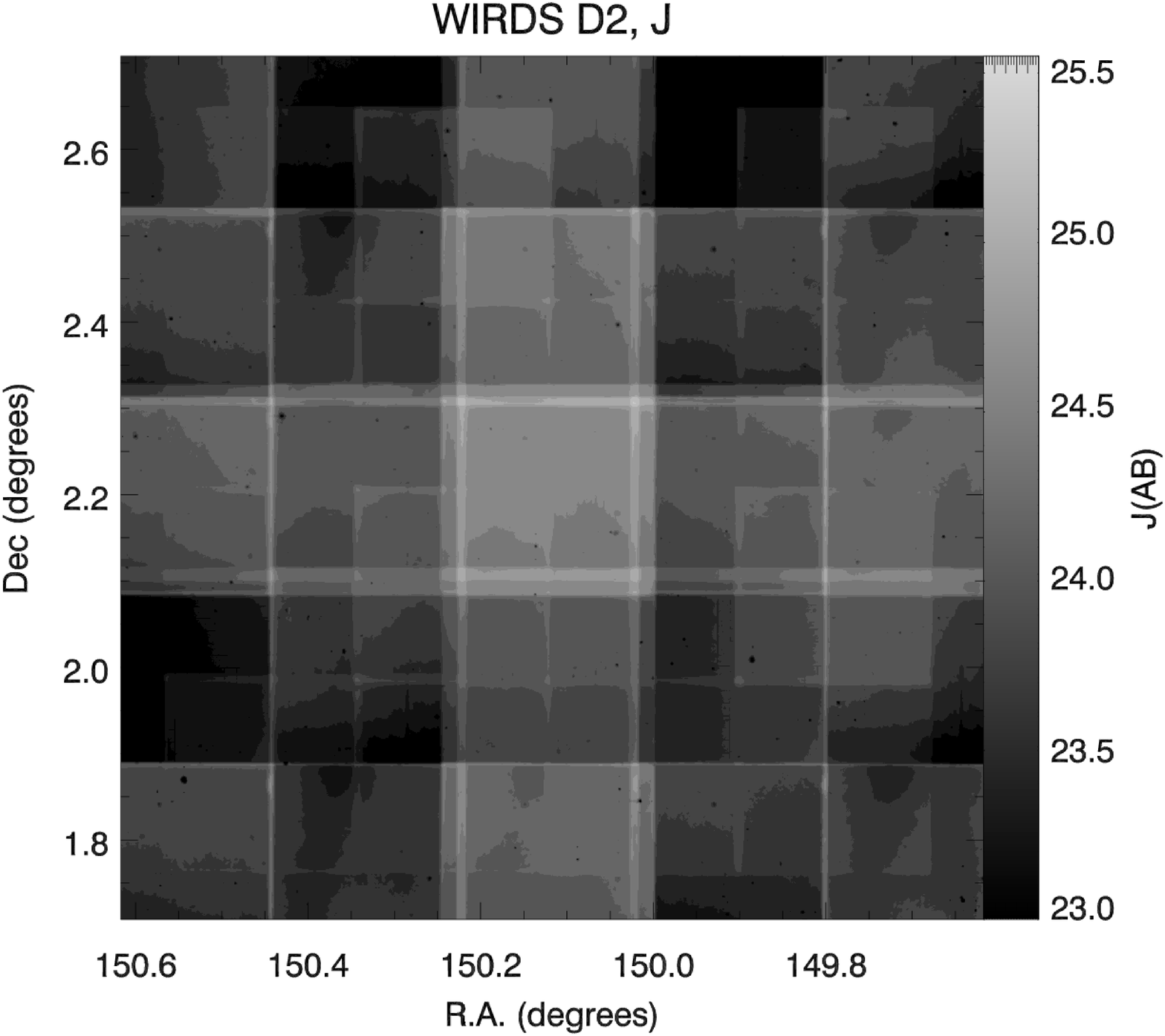}
\includegraphics[width=0.32\textwidth]{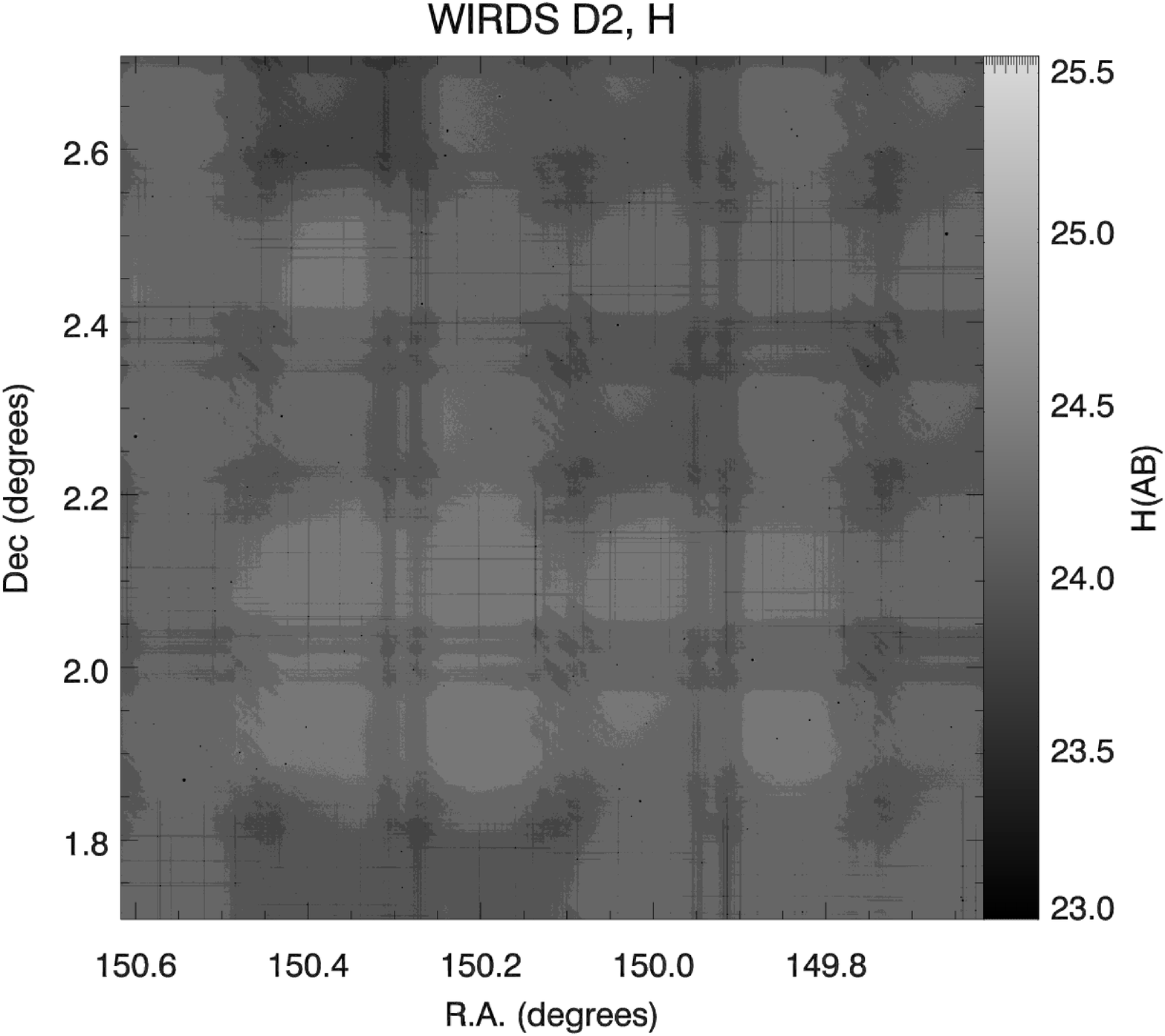}
\includegraphics[width=0.32\textwidth]{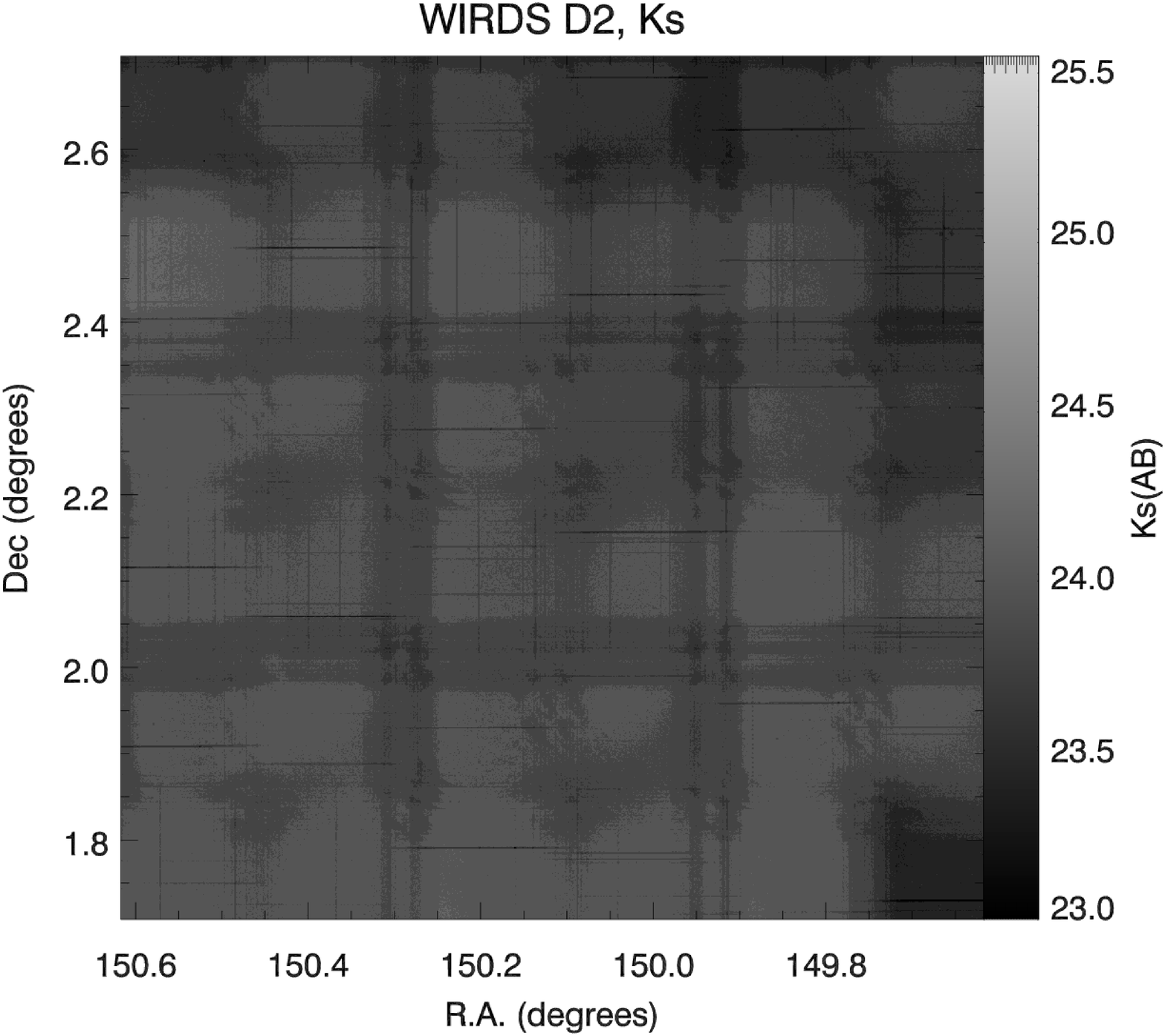}
   \caption{Weight maps for the D1 and D2 fields for each band. Each image covers $1^\circ\times1^\circ$; black areas are regions without data. The grey-scale gives the $3\sigma$ depth for 5 pixel sources in $2\arcsec$ apertures. Note that crosses of slightly shallower depth are present in each pointing due to the gaps between detectors (broad central crosses in each pointing) and also due to the use of single guide stars for many observations (narrow crosses present in individual quadrants in some pointings).}
   \label{fig:weights12}
\end{figure*}

\begin{figure*}
\includegraphics[width=0.32\textwidth]{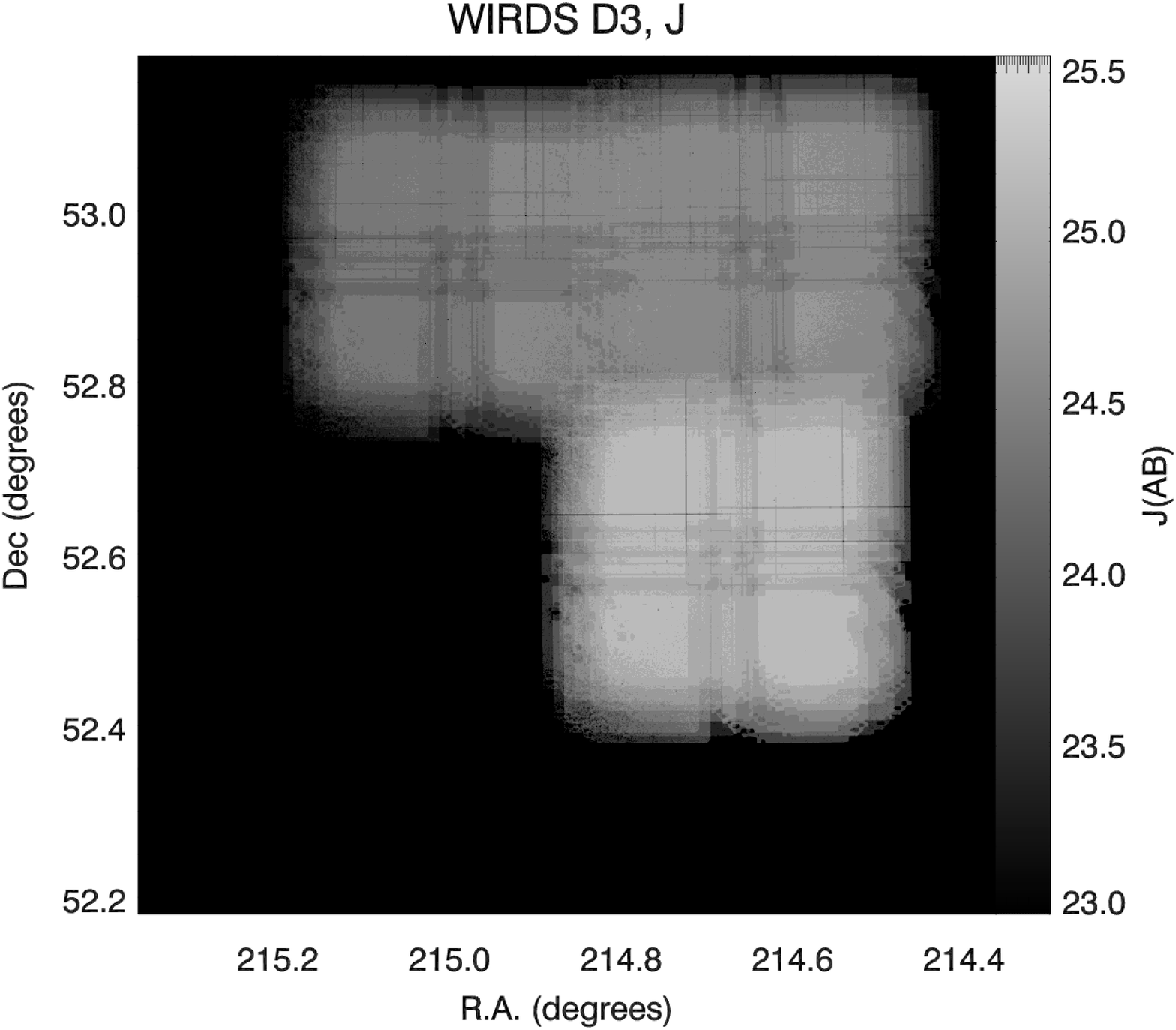}
\includegraphics[width=0.32\textwidth]{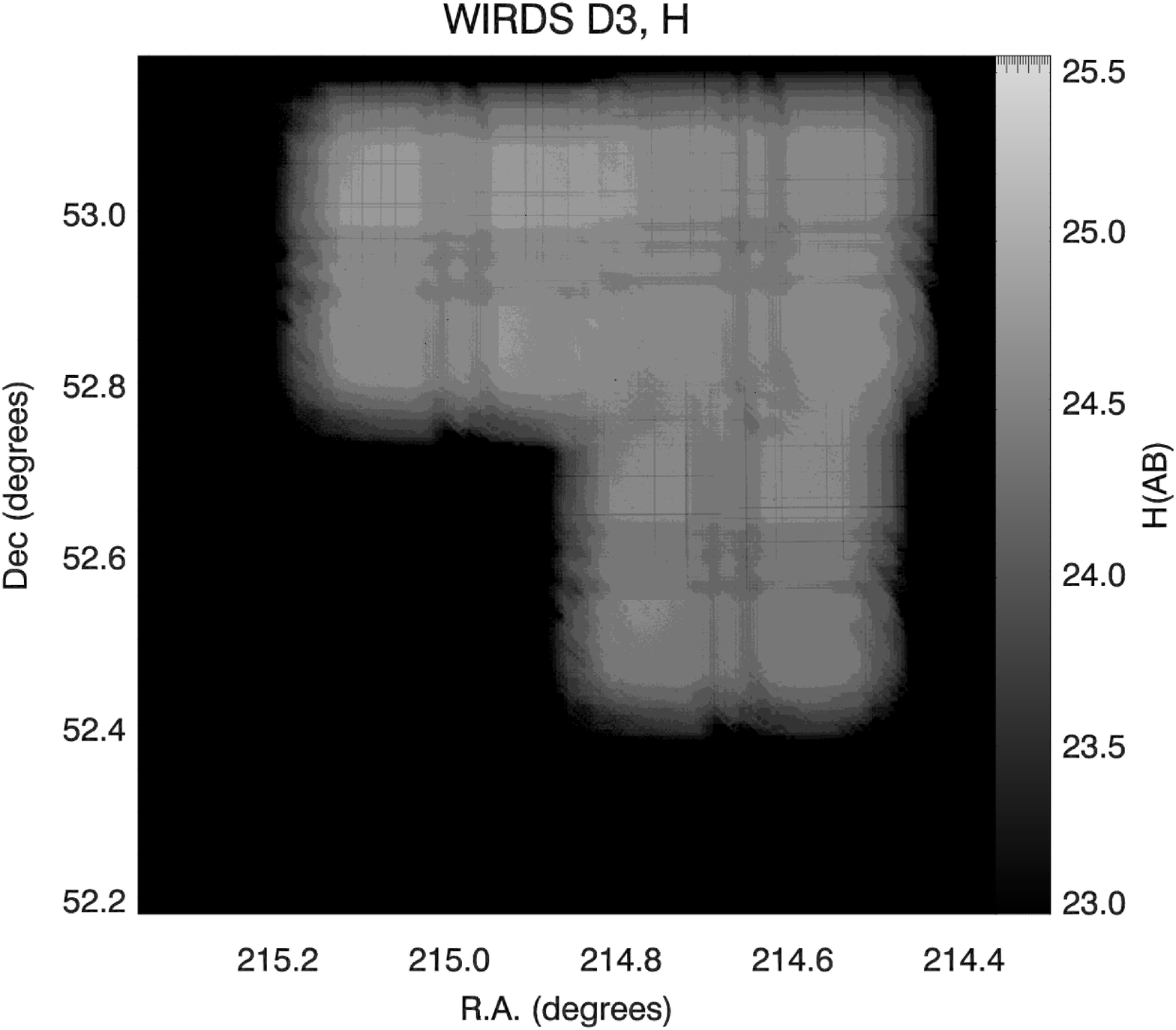}
\includegraphics[width=0.32\textwidth]{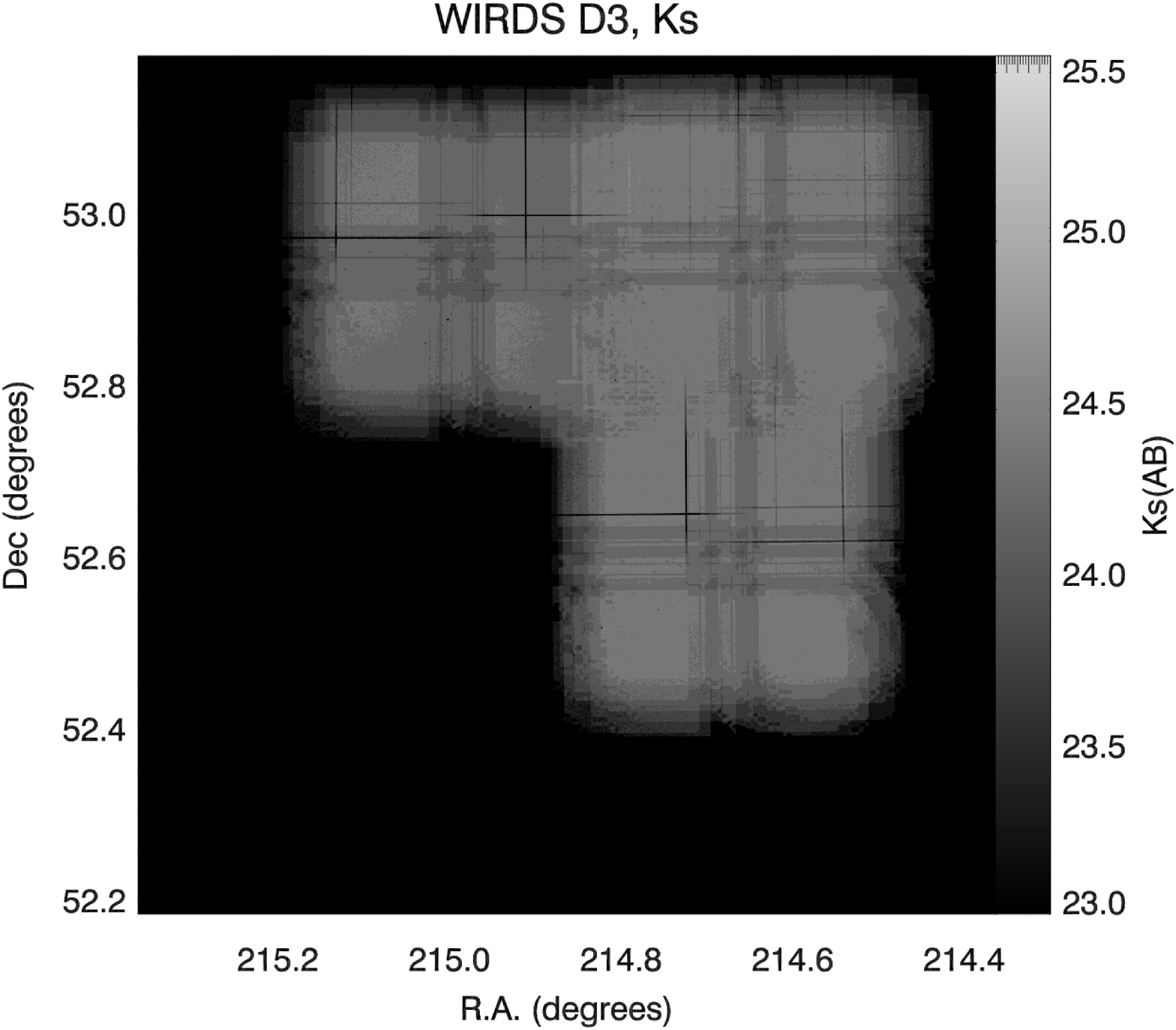}

\includegraphics[width=0.32\textwidth]{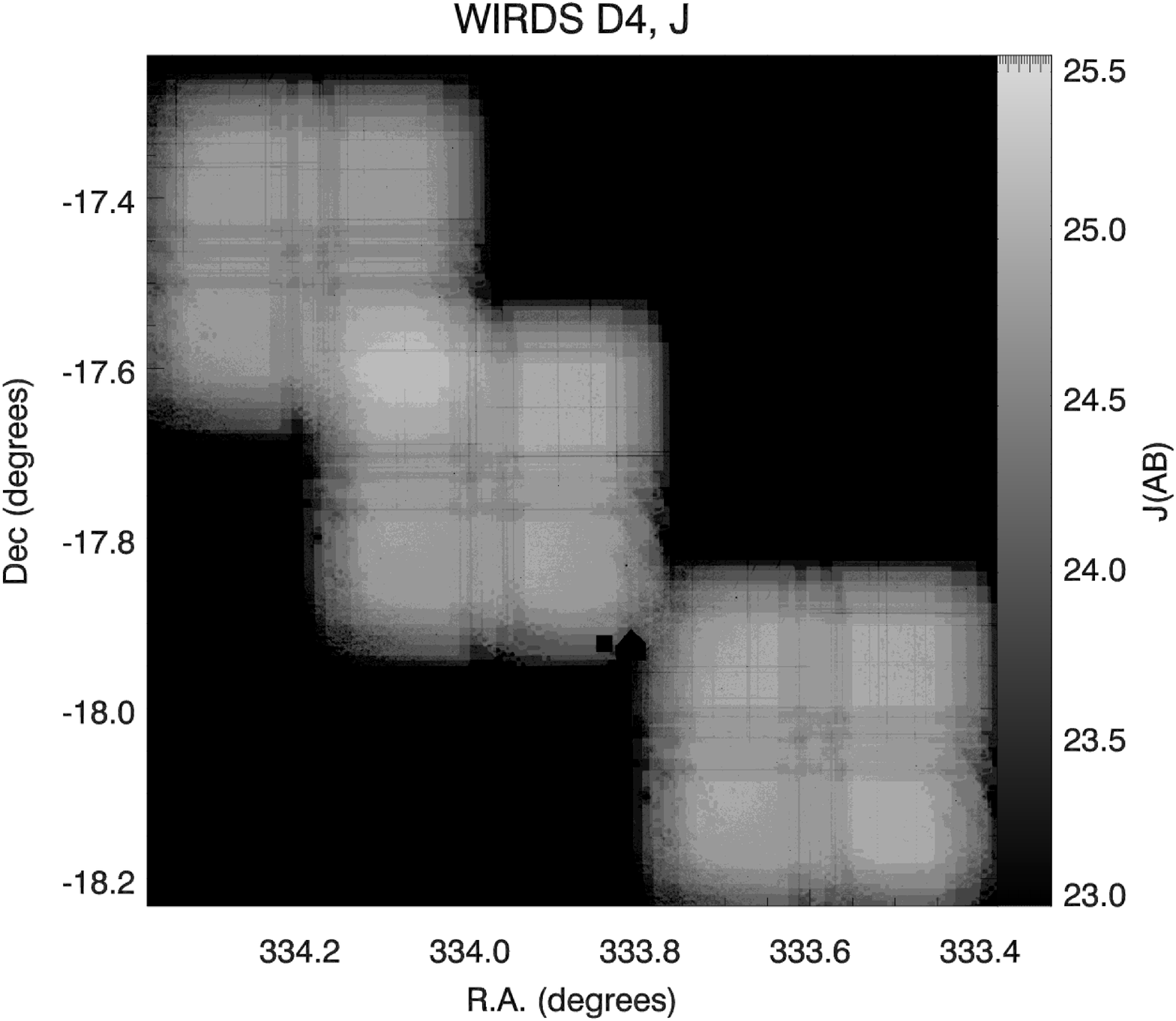}
\includegraphics[width=0.32\textwidth]{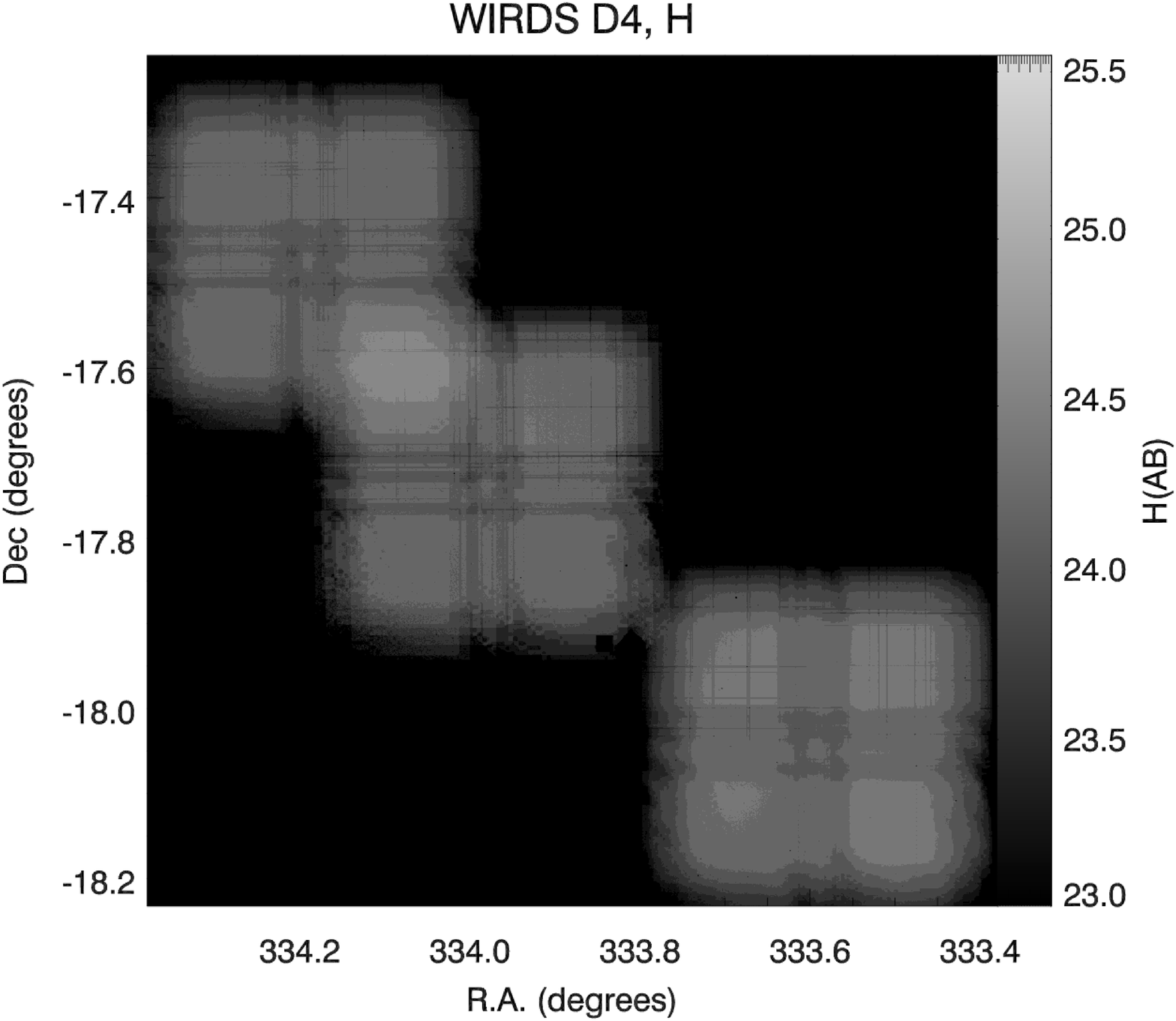}
\includegraphics[width=0.32\textwidth]{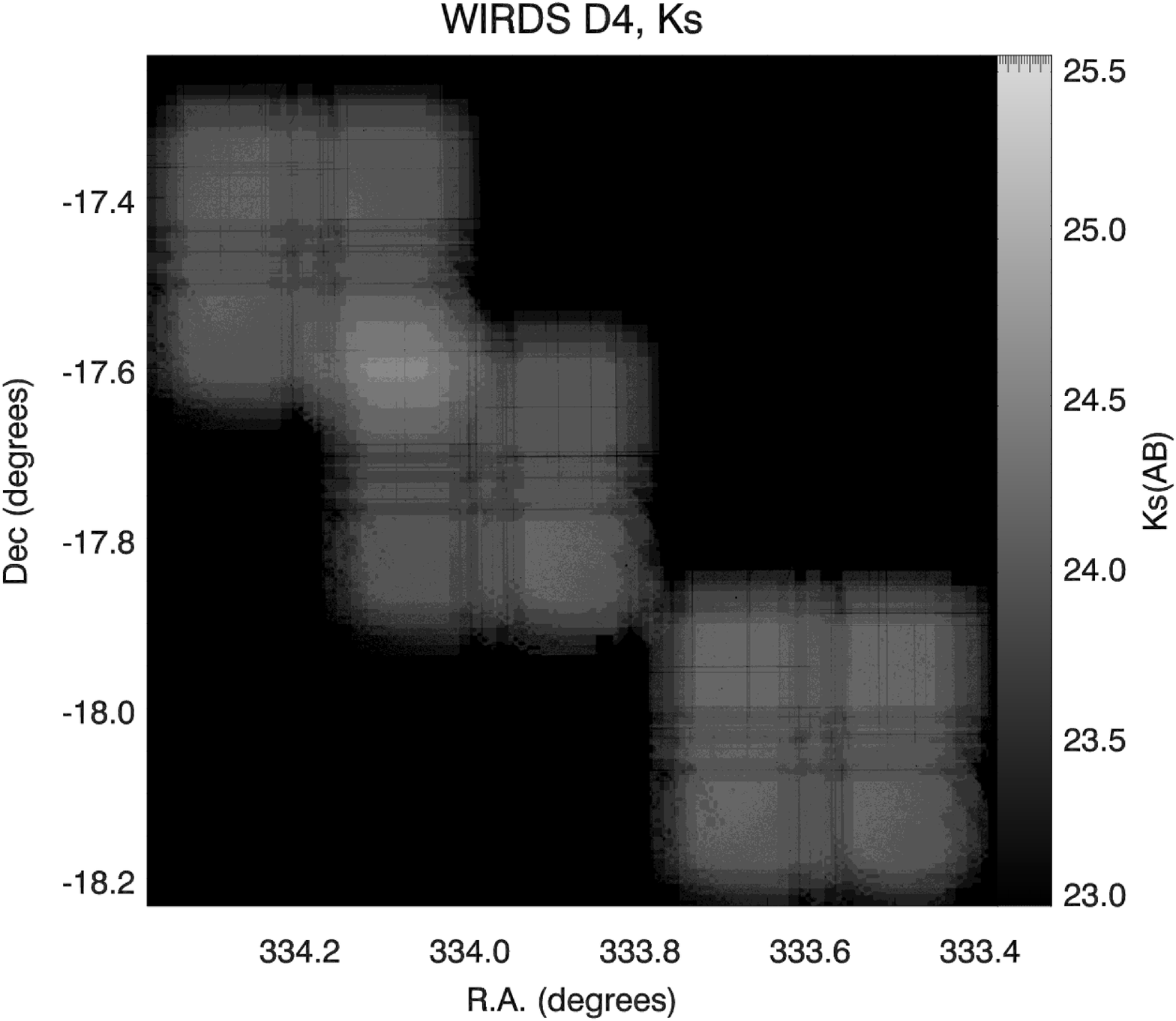}
   \caption{As in Fig.~\ref{fig:weights12}, but for the D3 and D4 fields.}
   \label{fig:weights34}
\end{figure*}

Each panel in Fig.~\ref{fig:weights12} (D1 and D2) and Fig.~\ref{fig:weights34} (D3 and D4) shows the weight maps for each field and  covers $1^\circ\times1^\circ$. The grey-scale represents the depth measured as the $3\sigma$ limit, based on a minimum object area of 5 pixels. This illustrates the excellent uniformity of the observations, especially in the final $H$ and $K_{\rm s}$ band stacks. We note that some depth variability between pointings can be seen in the $J$-band weight-maps in the D1 and D2 fields, which is due to loss of observing time through poor weather. Note that black regions in the thumbnails show areas lacking in any coverage. The COSMOS/D2 field is the only field with complete coverage of the entire CFHTLS Deep $1^\circ\times1^\circ$. The remaining three fields cover (prior to any masking) of $0.6\deg2$ (D1), $0.4\deg2$ (D3) and $0.4\deg2$ (D4).

We estimate the completeness of the final images by adding and detecting stellar sources in the stacks. Each image is split into individual pointings (5 for D1, 9 for D2 and 3 for D3 and D4) of $\sim20\arcmin\times20\arcmin$. The 50\% completeness limit for point-like objects is then estimated by placing simulated stellar-like objects at a random position for a given pointing with the appropriate PSF. We then use \texttt{SExtractor} to detect the simulated objects and measure the percentage of such objects successfully recovered based on 500 objects for a given magnitude interval.

\begin{figure*}
\centering
\includegraphics[width=\textwidth]{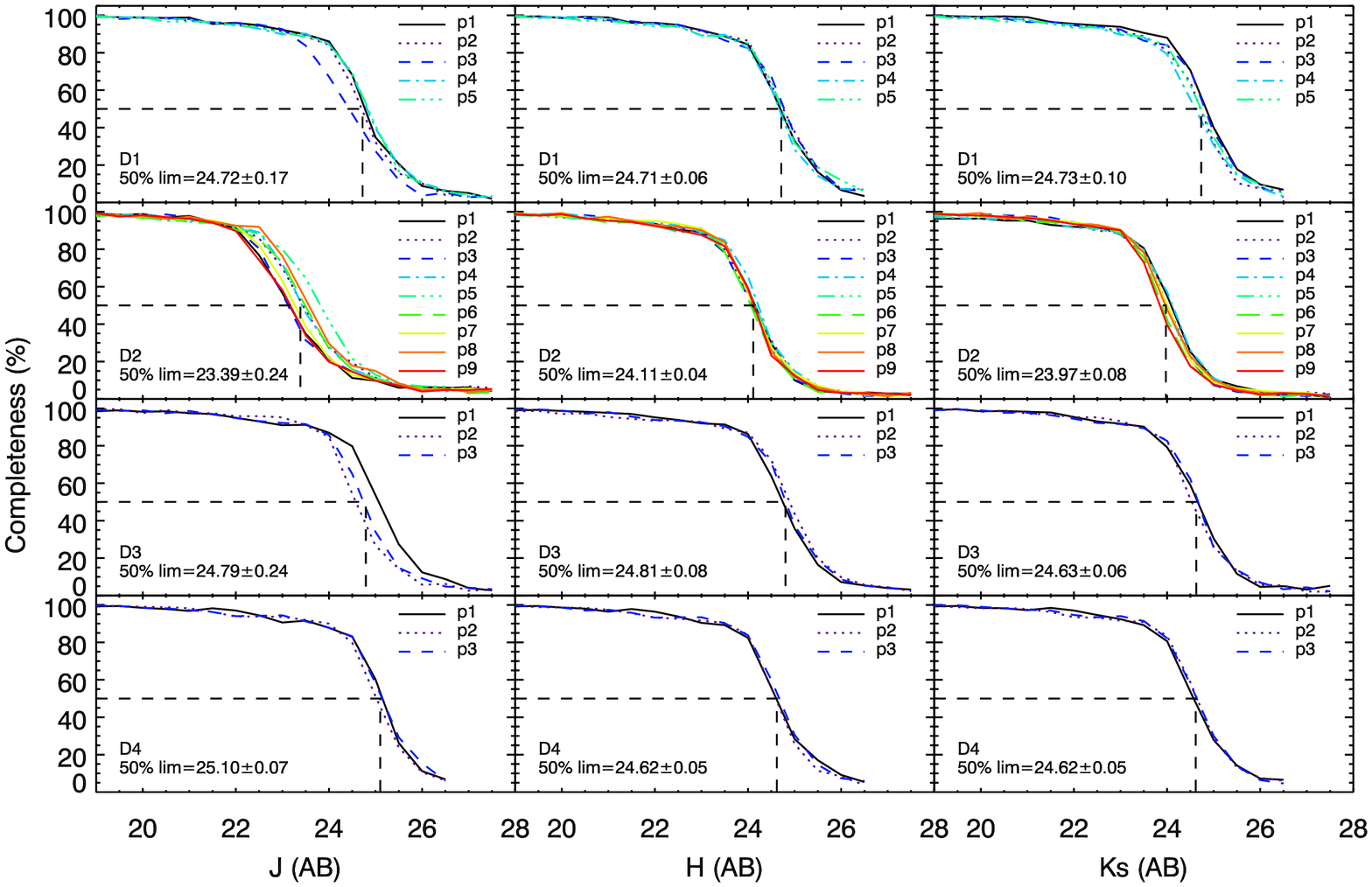}
   \caption{Point-source object completeness for the $J$ (left-hand panels), $H$ (middle panels) and $K_{\rm s}$ (right-hand panels) observations in the WIRDS fields. For each field/filter combination, the completeness is measured for each of the individual pointing regions (i.e. p1-p5 in the D1, p1-p9 in the D2, p1-p3 in the D3 and p1-p3 in the D4), illustrating the uniformity of the fields. The dashed lines in each case indicate the mean 50\% completeness for the given field/filter combination.}
   \label{fig:completeness}
\end{figure*}

Fig.~\ref{fig:completeness} shows our results for all 12 final images. Detection success rates are plotted for each individual pointing separately; it can be seen that our fields have similar completeness limits. In each case, we give the mean $50\%$ limit across all pointings for a given image (straight dashed lines in each plot) along with the standard deviation between those pointings. The 50\% completeness limits in the D1, D3 and D4 fields are consistently larger than 24.5AB in all three bands, whilst the D2 data reaches depths of $\approx24$AB in the $H$ and $K_{\rm s}$ bands and 23.4AB in the $J$-band image.

Consistent with the weight-maps, this figure shows that largest variation between pointings is for the D1, D2 and D3 $J-$ band images, however these are still consistent to $\sigma\sim0.2$ (as opposed to $\sigma\lesssim0.1$ for the $H$ and $K_{\rm s}$ images).

\subsection{Object extraction and photometry}

We have performed object extraction on the final NIR images in single and dual-image detection mode using \texttt{SExtractor} with a three pixel minimum detection area and a detection threshold of $1.5\sigma$. A background mesh size of 128 pixels was used for the background subtraction. Fluxes and magnitudes were calculated using the Kron \citep{1980ApJS...43..305K} -like total magnitudes (\texttt{MAG\_AUTO}) and $2\arcsec$ (10.75 pixels) and $4.5\arcsec$ (20 pixels) diameter aperture magnitudes. Note that for the \texttt{MAG\_AUTO} estimate, the minimum allowable aperture was set to $2.5\arcsec$. As seeing is relatively homogeneous across all the fields and bands and we do not perform any PSF homogenisation before catalogue extraction. 

\begin{figure}
\centering
\includegraphics[width=90.mm]{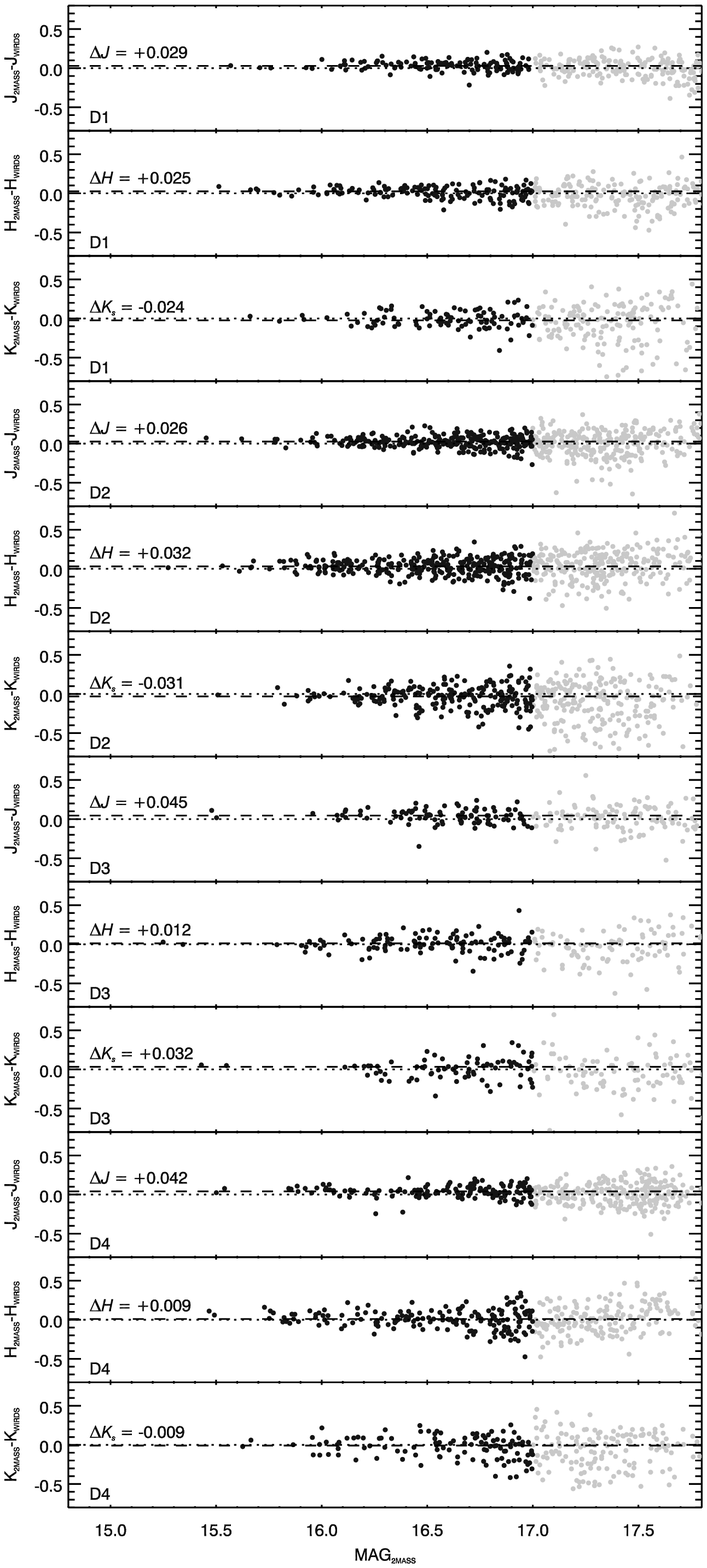}
   \caption{Photometric offsets for stars in WIRDS compared to 2MASS for each field and filter combination (note that these offsets have not been applied to the publicly released catalogues).}
   \label{fig:photometry}
\end{figure}

For dual-mode extraction, we used $gri$ ``$\chi^2$'' $K_{\rm s}$ band images as detection images, producing two sets of catalogues. In both cases we ran \texttt{SExtractor} in dual-mode with the detection image and all eight available images (i.e. $ugrizJHKs$) in each field from the CFHTLS T0006 Deep and WIRDS. The $gri$ $\chi^2$ was constructed from the CFHTLS T0006 Deep $gri$ images with \texttt{Swarp} in the ``CHI2'' combination mode\footnote{http://terapix.iap.fr/cplt/T0006-doc.pdf}. This produces an optimal combination of the input images where each pixel represents the probability of it being drawn from the sky distribution, based on a reduced $\chi^2$ of the background\citep{szalay99}. We compare the final photometry in our catalogues using 2MASS (which original individual exposures were calibrated, in Fig.~\ref{fig:photometry}. Small offsets are observed which for the most part are $\lesssim0.04$.

It is well known that \texttt{SExtractor} underestimates photometric errors for faint sources in combined images with correlated noise \citep[e.g.][]{2003AJ....125.1107L,2007ApJS..172..219L}. We estimate the magnitude of this effect by measuring the variance in blank apertures in random apertures. First a $4000\times4000$ subimage is extracted and aperture photometry is made with $2000\times2\arcsec$ apertures placed on blank regions of sky (note objects in the image are masked during this procedure). The standard deviation of the flux over all apertures is calculated. Next, we measure aperture photometry on all the SExtracted objects in the $4000\times4000$ pixel segment using $2\arcsec$ apertures. The correction factor is then taken as the ratio between the standard deviation of the blank sky apertures and the mean of the \texttt{SExtractor} errors for all the objects in that region. We repeat this process for a total of five $4000\times4000$ regions around each image and take the final correction to be the mean ratio across the image. This results in scaling of the magnitude errors of: $f_u=2.10$, $f_g=2.66$, $f_r=3.06$, $f_i=3.08$, $f_z=2.76$, $f_J=2.43$, $f_H=2.61$ and $f_{Ks}=2.49$. The final errors on the magnitudes are given by the \texttt{SExtractor} magnitude errors multiplied directly by $f_x$ (note that these corrections have not been applied to the publicly released catalogues). For the final (publicly released) catalogues, we do not apply any magnitude cut.

We estimate galactic dust-corrections for individual objects using the \citet{1998ApJ...500..525S} dust maps. Values of $A$ were estimated for each filter using the analytic formula of \citet{1989ApJ...345..245C}, giving $R(u)=4.705$, $R(g)=3.619$, $R(r)=2.679$, $R(i)=1.987$, $R(z)=1.522$, $R(J)=0.870$, $R(H)=0.571$ and $R(K_{\rm s})=0.366$.

In order to exclude regions where the photometry may be unreliable, for example due to bright stars or at the field edges, or unavailable, i.e. outside the NIR coverage, we apply masks to our final catalogues in order to flag objects. In each field, these masks are a combination of the CFHTLS optical masks and masks based on the WIRDS coverage. Once these masks have been applied to the catalogues, the total usable area is $0.49\deg2$ (D1), $0.8\deg2$ (D2), $0.4\deg2$ (D3) and $0.4\deg2$ (D4).


\subsection{Photometric redshifts and galaxy properties}

With our unique combination of optical and NIR data (with which we can detect the $4000\AA$ break to $z\sim4$) it is possible to estimate photometric redshifts and stellar masses reliably over a large redshift range. We use the ``Le Phare''\footnote{cfht.hawaii.edu/$\sim$arnouts/LEPHARE/cfht\_lephare} \citep{arnouts02,ilbert06} package to measure photometric redshifts and galaxy properties with a $\chi^2$ template-fitting method, based on the method of \citet{ilbert06} and \citet{2009ApJ...690.1236I}. As in \citet{2009ApJ...690.1236I}, we use a combination of the \citet{2007ApJ...663...81P} templates with the additional templates generated by \citet{2009ApJ...690.1236I} to account for starbursts with ages of 0.3 to 3 Gyr.  Following \citet{2010ApJ...709..644I}, we compute the final photometric redshifts using the median of the probability distribution function (PDFz).

\begin{figure}
\centering
\includegraphics[width=90.mm]{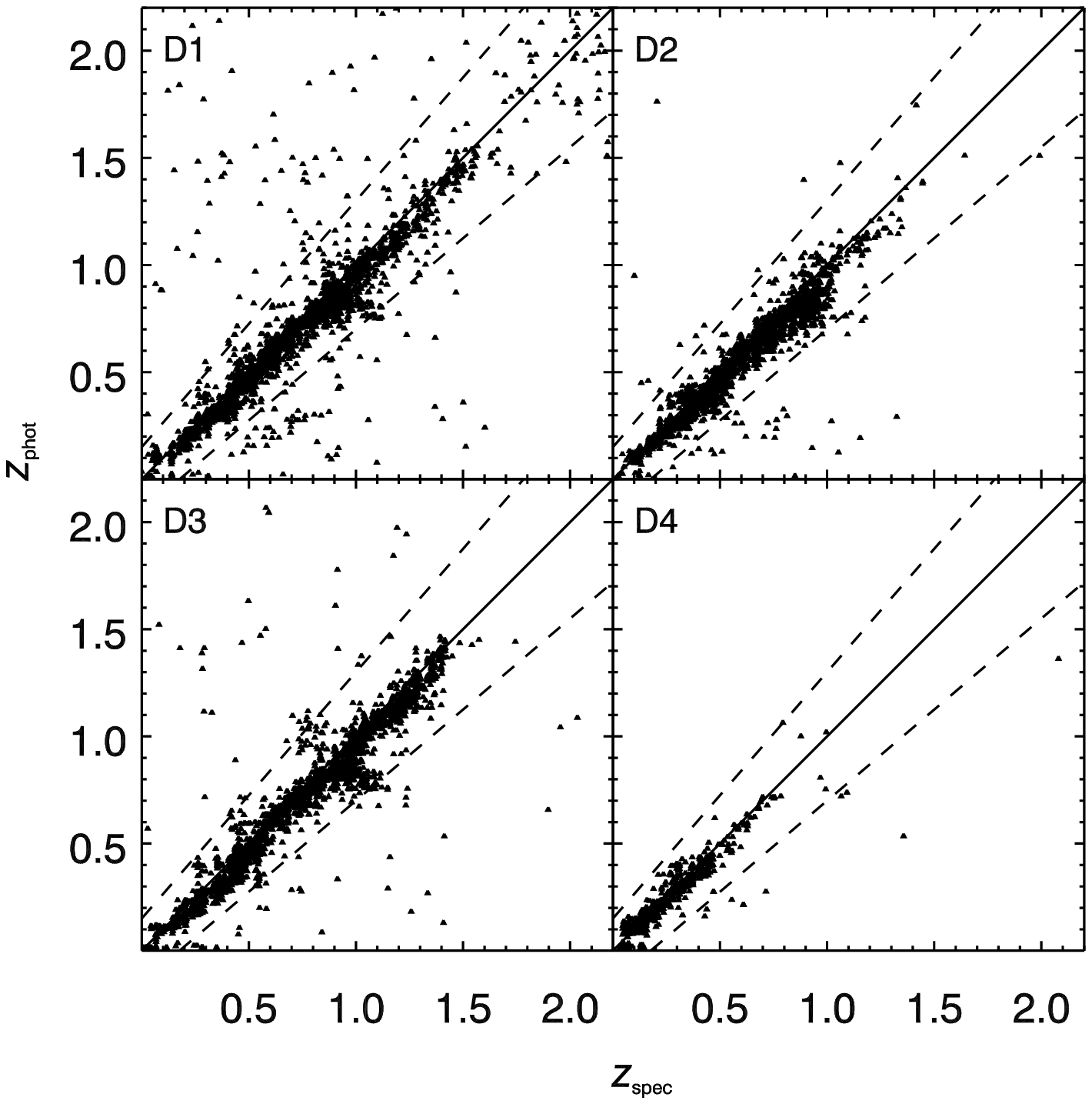}
   \caption{Photometric redshifts versus spectroscopic redshifts for all four fields. Top left shows the WIRDS D1 data with VVDS Deep and Ultra-Deep spectroscopic redshifts (3,192 galaxies). The top right panel shows the WIRDS D2 data compared to zCOSMOS 10K redshifts (3004 galaxies). Bottom left shows the WIRDS D3 data with DEEP2 spectroscopic data (2977 galaxies) and bottom right shows the WIRDS D4 data with the AAOmega spectroscopic redshifts (700 galaxies).}
   \label{fig:specVphot}
\end{figure}

We use an extensive collection of spectroscopic redshifts to calibrate our photometric redshifts. In D1, we use VVDS ``Deep'' \citep{2005A&A...439..845L} and ``Ultra Deep'' (\citealt{2012A&A...539A..31C}; Le F{\`e}vre et al. In Prep) spectroscopic samples. The publicly-available VVDS Deep sample contains 8,981 objects over $0.5\deg2$ in D1. It is a purely magnitude-limited sample with $i<24$ and spans a redshift range of $0\leq z\leq5$. The Ultra Deep sample comprises $\sim1500$ spectra over an area of $\approx0.14\deg2$ and covers a magnitude range of $22.5<i<24.75$. Both VVDS spectroscopic catalogues contain redshift quality flags which range from 1 to 4 with 1 being most unreliable and 4 being most reliable; in addition a flag 9 is given to objects identified based on a single emission line.  In this analysis we only use quality flags 3 and 4 from the VVDS catalogue (where these are the objects with the most secure spectroscopic classifications). Using the VVDS Ultra Deep data we find an outlier rate of $\eta=3.7\%$ and $\sigma_{\Delta z/(1+z)}=0.025$, with a median magnitude of $i^\ast_{\rm median}=24.0$, whilst using the Deep data we find an outlier rate of $\eta=1.9\%$ and $\sigma_{\Delta z/(1+z)}=0.030$, with a median magnitude of $i^\ast_{\rm median}=23.7$. The median offset between the photometric and spectroscopic redshifts is $\Delta z/(1+z)=0.017$.

\begin{figure}
\centering
\includegraphics[width=90.mm]{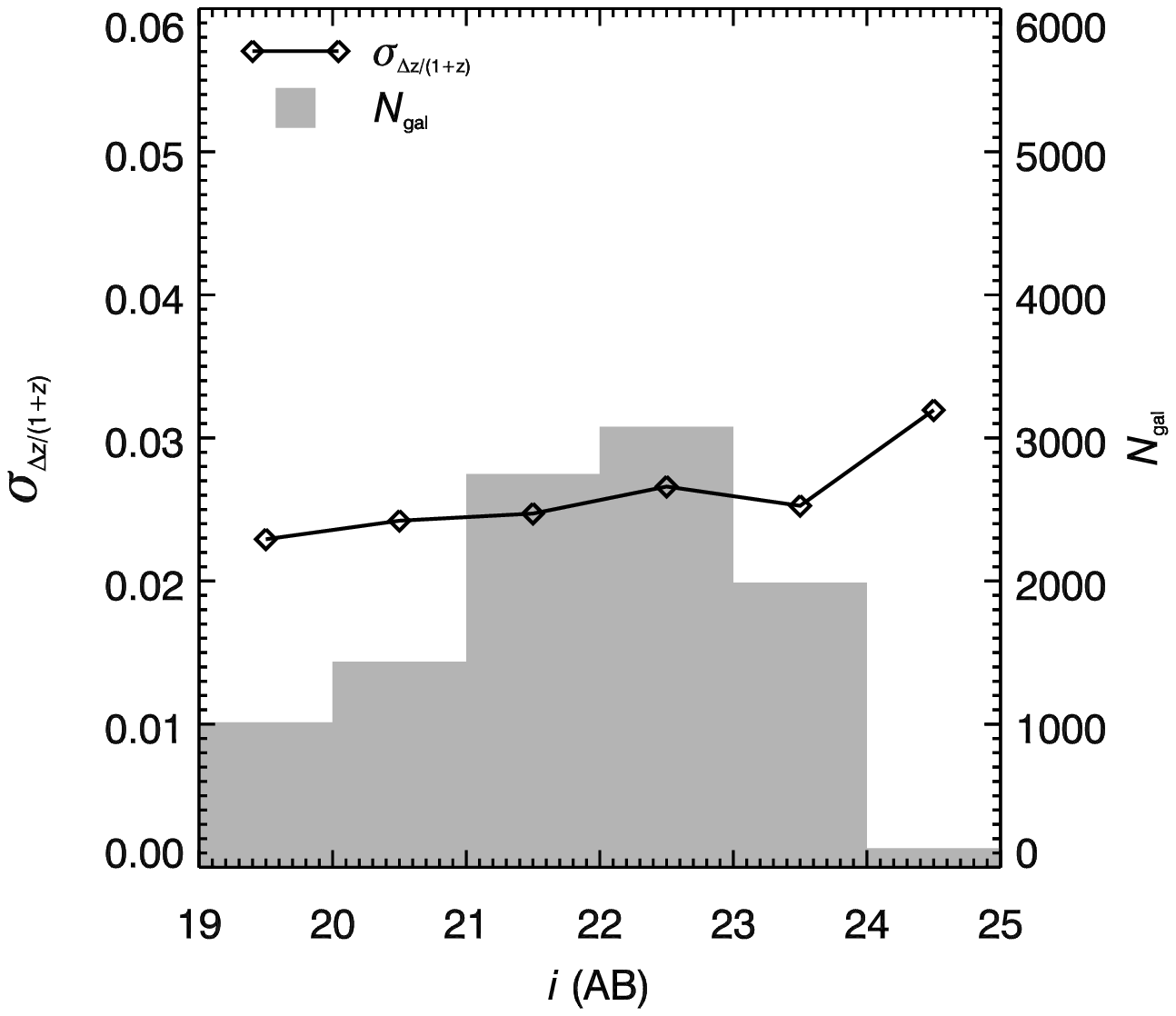}
   \caption{The redshift accuracy, $\sigma_{\Delta z/(1+z)}$, as a function of $i$-band magnitude for all four fields (connected diamond points). The spectroscopic data used in this analysis are described in the text, whilst the numbers of galaxies used for the comparison of spectroscopic versus photometric redshift at each magnitude is given by the grey histogram.}
   \label{fig:sigmaz}
\end{figure}

In the D2/COSMOS field we use the zCOSMOS ``10K'' spectroscopic sample \citep{2009ApJS..184..218L}. We find 3004 objects predominantly in the magnitude range $17.5<i<22.5$ and over a redshift range up to $z\lesssim1.4$ present in our photometric catalogue. From this data (and as before, using only objects with spectroscopic flags of 3 or 4) we estimate an outlier rate of $\eta=1.4\%$ and $\sigma_{\Delta z/(1+z)}=0.023$, based on a sample with median magnitude of $i^\ast_{\rm median}=21.6$ (spectroscopic observations for the VVDS and zCOZMOS in D1 and D2 fields were made using VIMOS on the VLT UT3). The median offset between the photometric and spectroscopic redshifts is $\Delta z/(1+z)=0.020$.

In D3, we use the DEEP2 DR3 redshift catalogue \citep{2003SPIE.4834..161D,2007ApJ...660L...1D}, based on observations using the Deep Imaging Multi-Object Spectrograph (DEIMOS) on Keck II. The catalogue contains 47,700 unique objects, of which 2,977 have a `zquality' flag of $\geq3$ (i.e. are deemed to be reliable redshifts) and are present in our photometric catalogue. This sample predominantly covers a magnitude range of $18<i<24$ with the bulk being below a redshift of $z\lesssim1.6$. Using the DEEP2 data, we estimate an outlier rate of $\eta=3.0\%$ and $\sigma_{\Delta z/(1+z)}=0.027$, based on a sample with median magnitude of $i^\ast_{\rm median}=22.5$ for the WIRDS D3 photometric catalogue. The median offset between the photometric and spectroscopic redshifts is $\Delta z/(1+z)=0.017$.

In the D4 field, we use spectra obtained using the AAOmega instrument on the Anglo-Australian Telescope (AAT) as part of a program to provide optical spectroscopy of X-ray point-sources in the CFHTLS \citep{2010MNRAS.401..294S}. The observations provide redshifts for 1,800 objects in the D4 field, of which 168 are QSOs, 66 are stars and 1,335 are galaxies, all at magnitudes of $i<22.5$ \citep{2010A&A...523A..66B}. In total, 1,090 of the galaxies overlap with our photometric data, most of which are at $z\lesssim0.8$. Based on these, we find an outlier rate of $\eta=2.1\%$ and $\sigma_{\Delta z/(1+z)}=0.021$, with a median magnitude of $i^\ast_{\rm median}=20.0$. The median offset between the photometric and spectroscopic redshifts is $\Delta z/(1+z)=0.013$. 

In Fig.~\ref{fig:specVphot} photometric redshifts are plotted against spectroscopic redshifts. The dashed line shows the outlier constraint of $\Delta z/(1+z)=0.15$. In addition, we plot the redshift accuracy, $\sigma_{\Delta z/(1+z)}$ as a function of $i$-band magnitude for all four fields combined in Fig.~\ref{fig:sigmaz}. These figures show that our photometric redshifts are highly accurate up to at least $z=1.5$ and $i=25$, with these limits imposed more by the available spectroscopic data than any reduction in photometric redshift accuracy. 

\subsection{Estimating stellar masses}

Following \citet{2010ApJ...709..644I} we use stellar population synthesis (SPS) models to estimate stellar masses effectively converting galaxy luminosity to a stellar mass \citep{2003ApJS..149..289B,2004A&A...424...23F}.

The SED templates were generated with the SPS package developed by \citet{bruzualcharlot03}.  Although these SPS models do not include the contribution of stars in the thermally pulsing asymptotic giant branch (TP-AGB) phase, we note that recent studies have shown the contribution of this phase to have been overestimated (by factors of $\sim3$) by models that do include it \citep{2010ApJ...722L..64K,2012arXiv1203.4571Z}. Significantly, \citet{2012arXiv1203.4571Z} find that the NIR SEDs of post-starburst galaxies (a population that would be expected to exhibit the strongest influence from the TP-AGB phase) are consistent with the SEDs predicted by the original \citet{bruzualcharlot03} models.

We assumed a universal \citet{2003PASP..115..763C} IMF and an exponentially declining star formation history of the form SFR $\propto e^{-t/\tau}$, with $\tau$ in the range 0.1 Gyr - 30 Gyr. The SEDs were generated for a grid of 51 ages in the range 0.1 Gyr - 14.5 Gyr. Dust extinction was applied to the templates using the \citet{2000ApJ...533..682C} law for $0<E(B-V)<0.5$. The templates were calculated based on two different values for the metallicity, $0.008Z_{\odot}$ and $0.02Z_{\odot}$. We impose a prior that significant extinction is only allowed for galaxies with high SFR, such that $E(B-V)<0.15$ for galaxy ages of $t>4\tau$.

\section{Galaxy populations and number counts}
\label{sec:bzk}

\subsection{$BzK$ photometric selections}

One of the key scientific objects of the WIRDS project is to probe the galaxy population at $z>1.4$, in particular via the ``$BzK$'' selection \citet{daddi04}, which is a means to isolate stars and galaxies of different types and redshifts. In the their original paper, \citet{daddi04} used $BzK$ colours to identify selection criteria to isolate galaxies at $1.4<z<2.5$ and to separate galaxies within this range into likely star-forming and passive galaxy samples. This has since been used by many authors to identify such galaxies and investigate their properties. At lower redshifts, \citet{2007MNRAS.379L..25L} and \citet{2010MNRAS.407.1212H} isolate features of representing passively evolving galaxies, with \citet{2007MNRAS.379L..25L} identifying ``branch" galaxies and \citet{2010MNRAS.407.1212H} identifying star-formation limits in the $BzK$ colour space. However, as pointed out by \citet{2011A&A...534A..81R}, colour selections can be potentially biased, with the effect of dust often significant. Using Spitzer 24 $\mu$m data, \citet{2011A&A...534A..81R} measured the success rate of the $BzK$ selection in identifying $1.4<z<2.5$ star-forming galaxies and found it to be $\sim90\%$ complete to their magnitude limit. Similarly, \citet{2007A&A...465..393G} presented an analysis of the $BzK$ selection, evaluating the limitations of the selection based on multi-wavelength data. They conlcude that the $BzK$ selection is highly efficient at identifying galaxies in the redshift range $1.4\leq z \leq 2.5$, but that for galaxies faint in the $K$ band and red in the $z-K$ colour, it is difficult to distinguish between star-forming and evolved galaxies, resulting in an underestimation of the passively evolving population.

\begin{figure}
\centering
\includegraphics[width=90.mm]{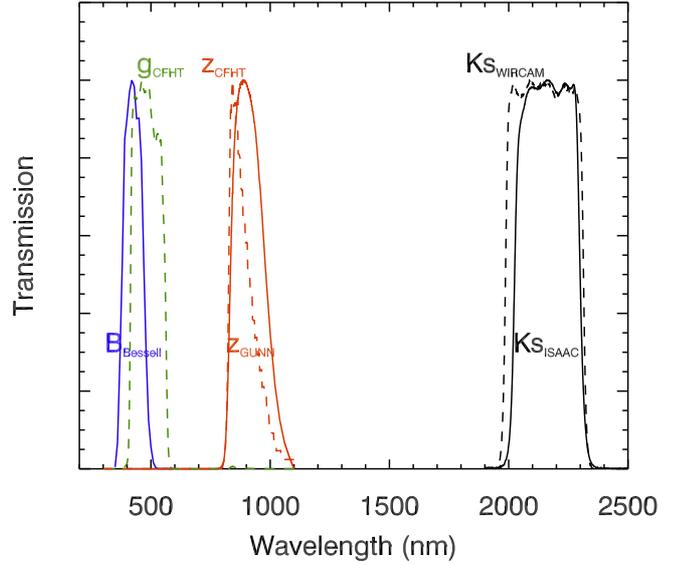}
   \caption{The $gzK$ and $BzK$ filter sets. The dashed lines show $g$ (green line) and $z$ (orange line) filters and the  $K_{\rm s}$ filter (black line). The solid lines show the Daddi et al, filters: $B_{Bessell}$ (blue line), $z_{GUNN}$ (orange-line) and the ISAAC $K_{\rm s}$ filter (black line).}
   \label{fig:filtset}
\end{figure}

In this Section we will present transformations from the $gzK$ photometry of CFHTLS/WIRDS to the $BzK$ colours of Daddi et al. and compare the selection results to that obtained based on our 8-band photometric redshift catalogue.

\begin{figure}
\centering
\includegraphics[width=90.mm]{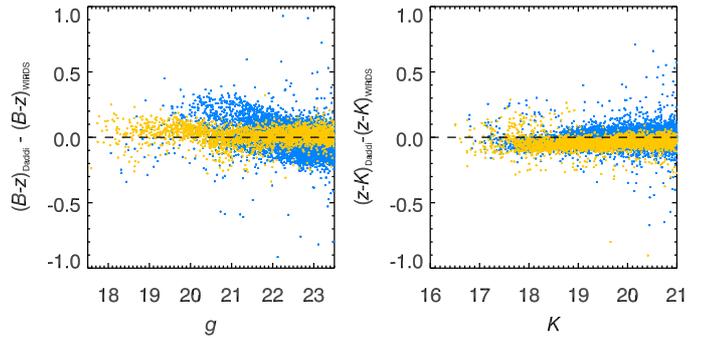}
   \caption{Comparison between COSMOS $BzK$ from \citet{2010ApJ...708..202M} and transformed WIRDS $gzK$ colours for objects common between the two datasets. Orange and cyan points are stars galaxies respectively.}
   \label{fig:gzk2bzk}
\end{figure}

In Figure~\ref{fig:filtset}, the solid blue, red and black curves show the $B$, $z$ and $K$ transmission profiles of the filters used by \citet{daddi04} and the dashed green, red and black curves show the MegaPrime $g$, $z$ and WIRCam $K_{\rm s}$ filters profiles. Although the $z$ and $K$ filters are similar, the MegaCam $g$ filter is offset from the Subaru $B$ filter.

We note that $BzK$ colours for objects the D2/COSMOS field has already been published \cite{2010ApJ...708..202M}. \citeauthor{2010ApJ...708..202M} make a transformation from their $B$, $z$ and $K$ filters to those of \citet{daddi04} colours using stellar spectra. However, the \citet{2010ApJ...708..202M} filters are overall closer to the original \citet{daddi04} filters. In particular, both works use the same $B$ filter. Given the differences between the $B$ and $g$ filters, the simple stellar spectra method cannot be used to transform between our system and the \citeauthor{daddi04} system. We therefore use the available COSMOS $BzK$ colours for galaxies and stars in the D2 field to compute an empirical transformation and to account for differences in galaxy spectral slopes. We first re-derive the colour transformation using a linear form for $(g-z)_{\rm WIRDS}$ to $(B-z)_{\rm Daddi}$ based purely on the stellar component and for the $(z-K)_{\rm WIRDS}$ to $(z-K)_{\rm Daddi}$ using both the stellar and galaxy components. We then correct the stellar transformation of the $(g-z)_{\rm WIRDS}$ to match the $(B-z)_{\rm Daddi}$ as a function of the $(u-g)_{\rm WIRDS}$ colours for the galactic component. The resulting offsets in the colours compared to the COSMOS data for both stars (orange points) and galaxies (blue points) are shown in Fig.~\ref{fig:gzk2bzk}. The transformations used to convert the CFHTLS/WIRDS colours to the $BzK$ colours of \citeauthor{daddi04} are then given by:

\begin{equation}
(B-z)_{\rm Daddi} = 0.410 + 1.217(g-z)_{\rm WIRDS} + C_{ug} \\
\end{equation}

\begin{equation}
(z-K)_{\rm Daddi} = 0.033 + 0.987(z-K)_{\rm WIRDS}
\end{equation}

\noindent where $C_{ug}$ is the correction based on the $(u-g)$ colour and is equal to:

\begin{equation}
C_{ug} = -0.318 + 0.291(u-g) 
\end{equation}

\noindent or equal to zero if the above is greater than zero.

\subsection{Galaxy number counts}

We now present the galaxy number counts for each of our fields. Galaxy counts are the basic and statistic galaxy population and provide an important ``zeroth-order'' test of any survey. Furthermore, near-infrared selected galaxy counts provide some are able to discriminate between certain types of galaxy evolutionary models \citep[e.g.][]{1993ApJ...415L...9G,1998ApJ...496L..93D,2000MNRAS.311..707M,2003ApJ...595...71C,2005MNRAS.361..701F,2006MNRAS.371.1601F,metcalfe06,2009MNRAS.398..497G,2009ApJ...696.1554C,2011MNRAS.414.1875H}.

In order to separate stars and galaxies we use the $BzK$ colour cut given by \citeauthor{daddi04}:

\begin{equation}
\label{eq:bzkstar}
(z-K_{\rm s}) > 0.3 (B-z) - 0.5
\end{equation}

Before applying this cut, we transform our colours to $BzK$ as described above and apply the photometric offsets given in Fig.~\ref{fig:photometry}. The resulting galaxy number counts in each of the three NIR bands are given in Table~\ref{table:galcounts}, combined across the four fields. In addition, we also give the standard deviation between the four fields for the counts in each band.

\begin{table*}
\caption{Differential number counts of galaxies in the $J$, $H$ and $K_{\rm s}$ bands.}             
\label{table:galcounts}      
\centering          
\begin{tabular}{l r r r r r r r r r r}    
\hline\hline       
  Mag (AB)&      $N_{gal}(J)$&      $N_{gal}(J)$& $\sigma_{F2F}(J)$&      $N_{gal}(H)$&      $N_{gal}(H)$& $\sigma_{F2F}(H)$&     $N_{gal}(Ks)$&     $N_{gal}(Ks)$& $\sigma_{F2F}(K_s)$\\
          &                & \multicolumn{2}{c}{(/0.5mag/deg$^{2})$}&                & \multicolumn{2}{c}{(/0.5mag/deg$^{2})$}&                & \multicolumn{2}{c}{(/0.5mag/deg$^{2})$}\\
\hline
     15.75&               5&                2.5&               1.2&              40&               19.7&               3.8&              49&               24.1&                 2.1\\
     16.25&              74&               36.4&               1.7&             107&               52.7&               5.3&             119&               58.6&                 6.0\\
     16.75&             149&               73.3&               7.9&             213&              104.8&              12.7&             252&              124.0&                12.5\\
     17.25&             296&              145.7&              14.6&             397&              195.4&              24.6&             478&              235.2&                29.9\\
     17.75&             491&              241.6&              27.8&             721&              354.8&              46.7&             848&              417.3&                43.4\\
     18.25&             849&              417.8&              46.4&            1169&              575.3&              72.7&            1445&              711.1&                89.4\\
     18.75&            1436&              706.7&              87.8&            2000&              984.3&             151.3&            2608&             1283.5&               166.3\\
     19.25&            2393&             1177.7&             145.0&            3347&             1647.1&             215.3&            4340&             2135.8&               252.8\\
     19.75&            3880&             1909.4&             224.0&            5282&             2599.4&             368.5&            6842&             3367.1&               411.1\\
     20.25&            6133&             3018.2&             354.0&            8057&             3965.1&             493.3&           10183&             5011.3&               589.3\\
     20.75&            8986&             4422.2&             466.6&           11846&             5829.7&             708.0&           14069&             6923.7&               796.8\\
     21.25&           12965&             6380.4&             613.2&           16407&             8074.3&             934.9&           19337&             9516.2&              1030.1\\
     21.75&           18634&             9170.3&             830.7&           22666&            11154.5&            1130.0&           26239&            12912.9&              1340.0\\
     22.25&           26433&            13008.4&            1260.6&           30417&            14969.0&            1437.5&           35093&            17270.2&              1771.0\\
     22.75&           36960&            18189.0&            1994.7&           38927&            19157.0&            1699.8&           44340&            21820.9&              2173.7\\
     23.25&           45692&            22486.2&            2456.2&           47804&            23525.6&            2133.9&           53912&            26531.5&              2419.8\\
     23.75&           48400&            23818.9&            1888.6&           50023&            24617.6&            2033.5&           58358&            28719.5&              1924.6\\
\hline
\end{tabular}
\end{table*}

\begin{figure}
\centering
\includegraphics[width=85.mm]{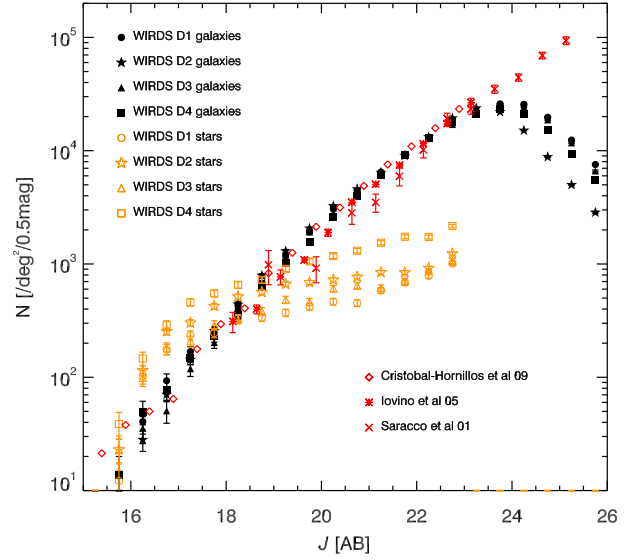}
   \caption{$J$ (AB) number counts from each of the WIRDS fields. The filled circles, stars, triangles and squares give the galaxy number counts for the D1, D2, D3 and D4 fields respectively. For comparison the red open upside down diamonds show the counts of \citet{2009ApJ...696.1554C}, the red asterisks show the \citet{iovino05} data and the red $\times$'s show the counts of \citet{saracco01}.}
   \label{fig:numcntJ}
\end{figure}

\begin{figure}
\centering
\includegraphics[width=85.mm]{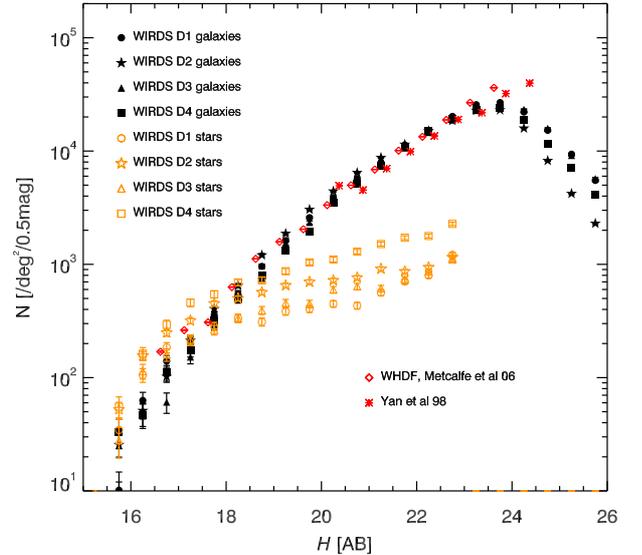}
   \caption{$H$ (AB) number counts from each of the WIRDS fields. The filled circles, stars, triangles and squares give the galaxy number counts for the D1, D2, D3 and D4 fields respectively. For comparison the red open diamonds show the counts of \citet{metcalfe06} and the red asterisks show the \citet{yan98} data.}
         \label{fig:numcntH}
\end{figure}

\begin{figure}
\centering
\includegraphics[width=85.mm]{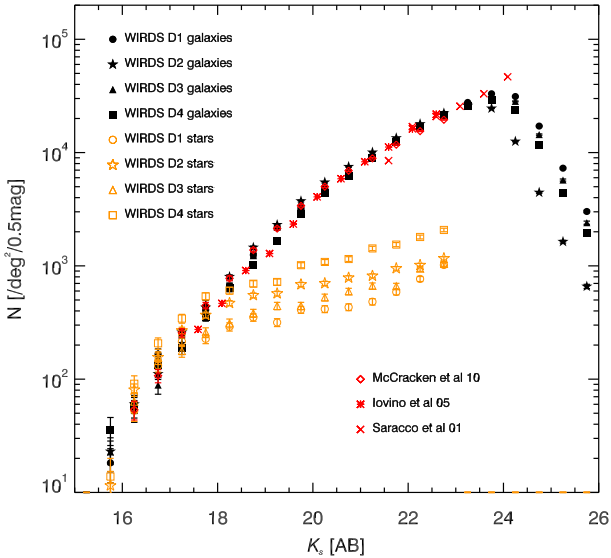}
   \caption{The $K_{\rm s}$ (AB) number counts from each of the WIRDS fields. The filled circles, stars, triangles and squares give the galaxy number counts for the D1, D2, D3 and D4 fields respectively. For comparison the red open diamonds show the counts of \citet{2010ApJ...708..202M}, the red asterisks show the \citet{iovino05} data and the red $\times$'s show the counts of \citet{saracco01}.}
   \label{fig:numcntKs}
\end{figure}

The number counts of galaxies in each of the four fields are shown in Fig.~\ref{fig:numcntJ} ($J$-band), Fig.~\ref{fig:numcntH} ($H$-band) and Fig.~\ref{fig:numcntKs} ($K_{\rm s}$-band). These are all measured directly from the individual $J$, $H$ or $K_{\rm s}$ selected catalogue in each case (as opposed $gri$- or $K_{\rm s}$- selected catalogues). In each case D1, D2, D3 and D4 counts are given by circles, stars, triangles and squares respectively. Both galaxy counts (filled black points) and stellar counts (open orange points) are shown for each field, with literature counts are plotted for comparison. For the $J$ band counts we plot counts from the VIRMOS Deep Imaging Survey \citep[][diamonds]{iovino05}, the ALHAMBRA survey \citep[][asterisks]{2009ApJ...696.1554C} and combined deep-NIR counts from the Chandra Deep and HDF South fields \citep[][$\times$'s]{saracco01}. In all four fields our number counts are consistent with those from the literature to $J\approx23$ beyond which our galaxy counts begin to be affected by incompleteness.

In $H$-band, we show the WHDF counts of \citet[][diamonds]{metcalfe06} and the HST/NICMOS observations of \citet[][asterisks]{yan98}. Again, close agreement is seen between the WIRDS number counts and the previously published data, with the completeness beginning to affect the WIRDS counts at $H\approx23.5$.

Finally, the $K_{\rm s}$ band counts are shown in comparison with the counts of \citet[][diamonds]{2010ApJ...708..202M} in the COSMOS field (of which our data includes a subset), the VIMOS Deep Imaging Survey counts of \citet[][asterisks]{iovino05} and the Chandra Deep and HDF South fields combined counts of \citet[][$\times$'s]{saracco01}.

Our stellar number counts show significant field-to-field variation, representing the different stellar populations in each deep field, with D4 having the highest density of stars and the D1 and D3 having lower densities. This is consistent with the galactic latitudes of the fields.

\subsection{Selection of star-forming and passive galaxies}

In this Section, we investigate a range of galaxy properties as function of type, i.e. star-forming versus passive galaxy populations. There are a number of ways to accomplish this, for instance by using observed colours, rest-frame colours or full template fitting. For the remainder of this paper, we will focus on using the derived rest-frame colours as a consistent method to separate star-forming and passive galaxy populations. For this, we follow \citet{2010ApJ...709..644I} and perform a selection in the derived rest frame colour $M_{\rm NUV}-M_r$\footnote{Here $M_{\rm NUV}$ is based on the GALEX rest-frame NUV filter and $M_r$ is based on the SDSS $r$ band filter.} (after dust correction is performed). In \citet{2010ApJ...709..644I}, three selection criteria are used to identify star-forming, passive and intermediary populations. Here we simplify this to two populations, star-forming (incorporating the star-forming and intermediary populations of \citealt{2010ApJ...709..644I}) and passive. Thus the selection criteria take the following form:

\begin{equation}
M_{\rm NUV}-M_r \ge 3.5\rm{~~(Passive)} 
\end{equation}
\begin{equation}
M_{\rm NUV}-M_r < 3.5\rm{~~(Star-forming)}
\end{equation}

The separation of star-forming versus passive populations in this way is proven to be a good indication of star-formation activity \citep[e.g.][]{2007ApJS..173..342M,2007A&A...476..137A} and its success has been highlighted by the comparison with morphological classifications in \citet{2010ApJ...709..644I}.

\subsection{Galaxies at $z<1.4$}

\begin{figure}
\centering
\includegraphics[width=90.mm]{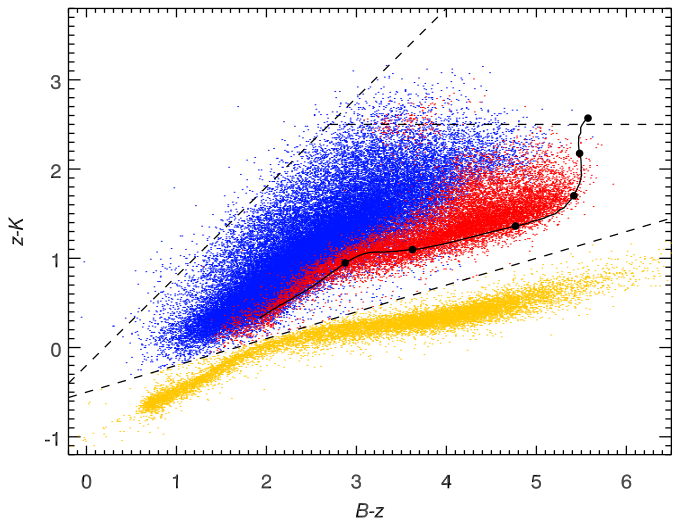}
   \caption{The $BzK$ diagram for galaxies at $z<1.4$. The galaxy population is split into two groups: red points for passive galaxies, and blue points for star-forming galaxies (separated using rest-frame $M_{\rm NUV}-M_r$ colours). Only objects with $K_{\rm s}<22$ and a photometric redshift of $z<1.4$ are shown. The black dashed lines show the cuts used by \citet{daddi04} to select $z>1.4$ galaxies and stars, whilst the solid black line shows the model track for a passively evolving elliptical galaxy from $z=1.5$ to $z=0$. The filled black circles on this line are at intervals of $\Delta z=0.25$, beginning at $z=0.25$ and finishing at $z=1.5$.}
   \label{fig:bzk_lowz}
\end{figure}

We first look at the properties of the $z<1.4$ galaxy population in the WIRDS data, making use of the full $gri$ $\chi^2$ selected photometric catalogue. Fig.~\ref{fig:bzk_lowz} shows the $BzK$ diagram for all stars and galaxies in WIRDS with photometric redshifts of $z<1.4$ and a magnitude limit of $K_{\rm s}<22$. Orange points show the stellar population (identified using the $BzK$ cut), red the passive galaxies and blue points show the star-forming galaxy population (identified as described above using the rest-frame $M_{\rm NUV}-M_r$ colours of the galaxies). The dashed lines show the $BzK$ colour cuts presented by \citet{daddi04}, which are given by equation~\ref{eq:bzkstar} for the stellar-galaxy separation, whilst the selection of star-forming galaxies at $z>1.4$ is given by:

\begin{equation}
\label{eq:sbzk}
(z-K_{\rm s}) > (B-z)-0.2
\end{equation}

\noindent and the boundary for passive galaxies at $z>1.4$ is given by:

\begin{equation}
\label{eq:pbzk}
(z-K_{\rm s}) < (B-z)-0.2 \cap (z-K_{\rm s}) > 2.5
\end{equation}

These selection criteria follow closely the boundaries of our $BzK$ transformed colour selection and photometric redshift limits to a good degree, with few $z<1.4$ galaxies lying outside the $z>1.4$ criteria. Based on our photometric redshifts, we find that only 0.5\% and 0.5\% of all $K<24$, $z<1.4$ galaxies fall within the $z>1.4$ star-forming and passive galaxy selections respectively. The stellar locus also appears to be well separated by the \citet{daddi04} criteria, suggesting that our $gzK$ to $BzK$ transformation are correct. As discussed by \citet{2010MNRAS.407.1212H}, there is a separation of star-forming and passive galaxies in the $BzK$ at $z<1.4$ as well as at $z>1.4$. The passive galaxies in fact form a locus approximately parallel to the stellar locus at $B-z\gtrsim3$, already noted by \citet{2007MNRAS.379L..25L} in UKIDSS data. We highlighted this in Fig.~\ref{fig:bzk_lowz} with the evolutionary track (black line with filled circles at intervals of $\Delta z=0.25$) of a typical passively evolving galaxy from $z=1.5$ to $z=0$, calculated using the galaxy evolution code of \citet{bruzualcharlot03}, a Salpeter IMF and an instantaneous starburst and a redshift of formation of $z_f=5$.

\subsection{Passive and star-forming galaxies at $z>1.4$}

\begin{figure}
\centering
\includegraphics[width=90.mm]{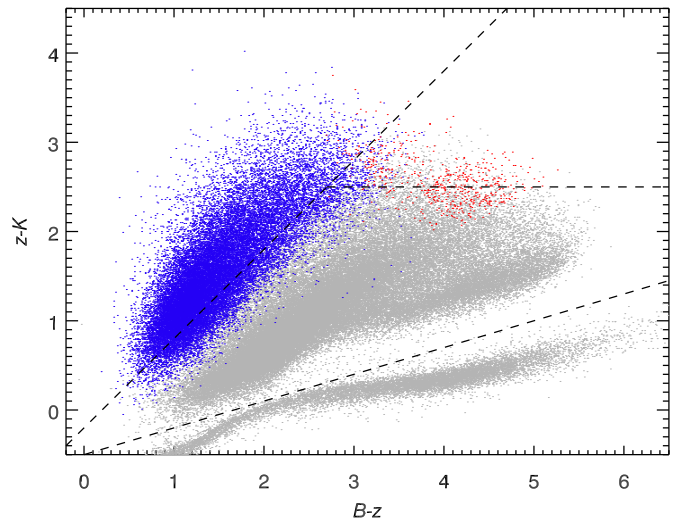}
   \caption{The $BzK$ diagram for $z>1.4$ galaxies. The blue and red points show star-forming and passive galaxies at $z>1.4$ (based on the $M_{\rm NUV}-M_r$ rest-frame colours and photometric redshifts). The grey points show the stellar and $z<1.4$ galaxy populations (as plotted in Fig.~\ref{fig:bzk_lowz}). The dashed lines show the selection criteria of \citet{daddi04}.}
   \label{fig:bzk_hiz}
\end{figure}

We now turn to galaxies at $z>1.4$, plotting once again the $BzK$ diagram for the WIRDS data in Fig.~\ref{fig:bzk_hiz}. All points from Fig.~\ref{fig:bzk_lowz} are plotted with grey points (i.e. stars and $z<1.4$ galaxies). Overplotted on this $z<1.4$ population are galaxies with $z>1.4$ with blue points showing star-forming galaxies and red points showing passive galaxies. Again the dashed lines show the colour constraints of \citet{daddi04} as given in equations~\ref{eq:bzkstar}, \ref{eq:sbzk} and \ref{eq:pbzk}. The s$BzK$ constraint corresponds closely to galaxies with $z_{\rm phot}>1.4$ and with high star-formation rate. At $K<24$, 23\% of objects are scattered into the $z<1.4$ region, however the bulk (76\%) are at $z>1.4$. The final 1\% of star-forming objects (based on the rest-frame colour selection) are found to fall within the $z>1.4$ passive galaxy selection. We find a somewhat lower success rate than \citet{2011A&A...534A..81R}, who found a success rate of 90\%. This is probably a consequence of the increased scatter in our photometric redshift estimates in comparison with their work, which includes many more photometric bands in addition to, crucially, Spitzer 24 $\mu$m measurements in their photometric redshifts. However, both results ultimately show the effectiveness of the $BzK$ selection in identifying star-forming galaxies at $z>1.4$. 

\begin{figure}
\centering
\includegraphics[width=90.mm]{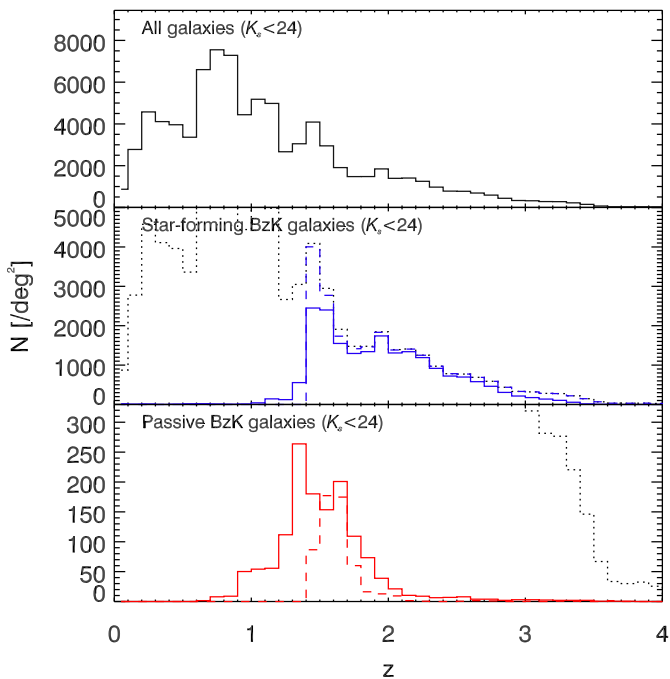}
   \caption{\emph{Top:} Photometric redshift distributions of the entire galaxy population at $K_{\rm s}<24$. \emph{Middle:} Redshift distribution of galaxies selected to be star-forming at $z\gtrsim1.4$ via the s$BzK$ selection (solid blue histogram) and using photometric redshifts and template fits (dashed blue histogram). The dotted line shows the entire $K_{\rm s}$ population. \emph{Bottom:} Redshift distribution of galaxies selected to be passive at $z\gtrsim1.4$ via the p$BzK$ selection (solid red histogram) and again using photometric redshifts and template fits (dashed red histogram).}
   \label{fig:bzk_nz}
\end{figure}

Concerning the passive $z>1.4$ population, passive objects selected from the photometric redshift catalogue significantly overlap with the $z<1.4$ population. Many passive $z>1.4$ galaxies are missed by the p$BzK$ cut when applied to our data, falling in either the $z<1.4$ region (41\%) or the star-forming $z>1.4$ region (7\%). In addition, the p$BzK$ selection includes a significant fraction of $z<1.4$ objects. We note that the p$BzK$ selection relies on extremely deep $B$ (or in this case $g$) band data and given the $g$ band depths of $\approx26$, these numbers for the p$BzK$ selection are likely affected by incompleteness and increased photometric errors compared to deeper datasets. Despite this, we note that similar results have been published, with \citet{2007A&A...465..393G} for example reporting a completeness of only $34\%$ for the p$BzK$ selection in terms of retrieving the passive $1.4<z<2.5$ galaxy population.

Fig.~\ref{fig:bzk_nz} shows the photometric redshift distribution for at $K_{\rm s}<24$ for p$BzK$ and s$BzK$ selected galaxy populations. The top panel shows the entire $K_{\rm s}<24$ galaxy redshift distribution (solid black line). This extends well out to $z\approx3.5-4$. 

The middle panel shows the s$BzK$ galaxy redshift distribution (solid blue histogram). We also show the $z>1.4$ star-forming population as selected using the photometric redshift catalogue in which star-forming galaxies are identified via the rest-frame colours (dashed blue histogram). Comparing the two, the photometric redshifts suggest that the s$BzK$ selection misses a small fraction of the $z\approx1.4$ galaxy population, whilst including a small number of galaxies in the range $1<z<1.4$.

In the bottom panel of Fig.~\ref{fig:bzk_nz}, we show the p$BzK$ redshift distribution (solid red histogram). We also show the $z>1.4$ passive population identified in the template fitting using Le Phare (dashed red histogram). In agreement with the $BzK$ distribution we observe in Fig.~\ref{fig:bzk_hiz}, we see a significant contribution to the p$BzK$ selection from galaxies identified with photometric redshifts of $1<z<1.4$.

In form, the $BzK$ redshift distributions are similar to those measured by \citet{2005ApJ...633..748R}, \citet{2007A&A...465..393G}, \citet{2008MNRAS.391.1301H} and \citet{2010ApJ...708..202M}, with the s$BzK$ selection showing a relatively sharp lower-redshift cut-off at $z=1.4$ and the p$BzK$ selection extending to lower redshifts of $z\approx1$. Additionally, the s$BzK$ selection extends to relatively large redshifts of $z>3$. In contrast, few p$BzK$ galaxies are observed with photometric redshifts of $z\gtrsim2$.

\begin{figure}
\centering
\includegraphics[width=90.mm]{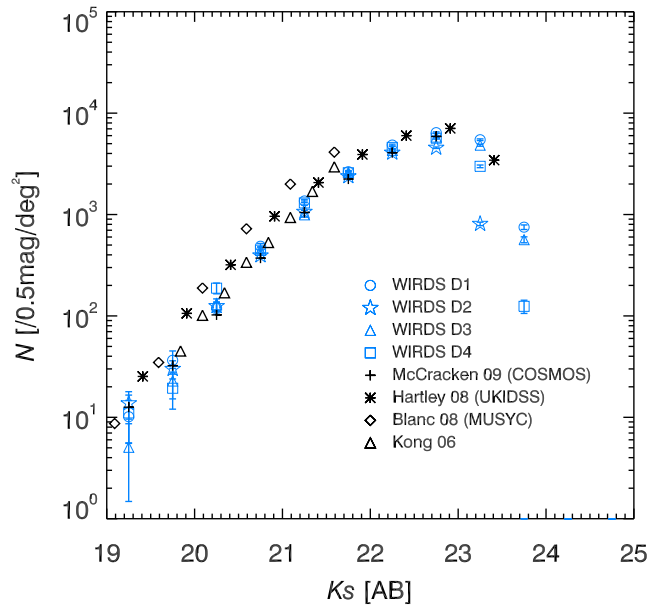}
   \caption{Number counts of galaxies selected using the s$BzK$ colour selection. Blue circles, stars, triangles and squares show the number counts of s$BzK$ galaxies in the WIRDS D1, D2, D3 and D4 fields respectively. Black points show comparison counts from the literature, with the triangles showing the counts of \citet{2006ApJ...638...72K}, diamonds the counts of \citet{blanc08}, asterisks those of \citet{2008MNRAS.391.1301H} and crosses those of \citet{2010ApJ...708..202M}.}
   \label{fig:numcntsBzK}
\end{figure}

We show number counts for s$BzK$ samples in Fig.~\ref{fig:numcntsBzK}. Also shown are counts of \citet[][triangles]{2006ApJ...638...72K}, \citet[][diamonds]{blanc08} from the MUSYC Survey, \citet[][asterisks]{2008MNRAS.391.1301H} from UKIDSS and \citet[][crosses]{2010ApJ...708..202M} from COSMOS. In general we see good agreement between our counts of s$BzK$ galaxies and the literature counts, although the counts of \citet{2008MNRAS.391.1301H} and in particular \citet{blanc08} are marginally higher than the WIRDS counts. The s$BzK$ number counts combined across all four WIRDS fields are given in Table~\ref{table:sbzkcounts}.

\begin{table}
\caption{WIRDS s$BzK$ number counts in the four WIRDS field}             
\label{table:sbzkcounts}      
\centering          
\begin{tabular}{l r r}    
\hline\hline       
$K_{\rm s}$(AB) & $N_{\rm gal}$&$ N_{\rm gal}$\\
(AB) & &  (/0.5mag/deg$^{2})$  \\
\hline                    
18.25 &       3 &       1.5 \\
18.75 &       6 &       3.0 \\
19.25 &      22 &      10.8 \\
19.75 &      58 &      28.5 \\
20.25 &     278 &     136.8 \\
20.75 &     892 &     439.0 \\
21.25 &    2385 &    1173.7 \\
21.75 &    5169 &    2543.8 \\
22.25 &    9007 &    4432.6 \\
22.75 &   11155 &    5489.7 \\
23.25 &    6332 &    3116.1 \\
23.75 &     638 &     314.0 \\
\hline                  
\end{tabular}
\end{table}

\begin{figure}
\centering
\includegraphics[width=90.mm]{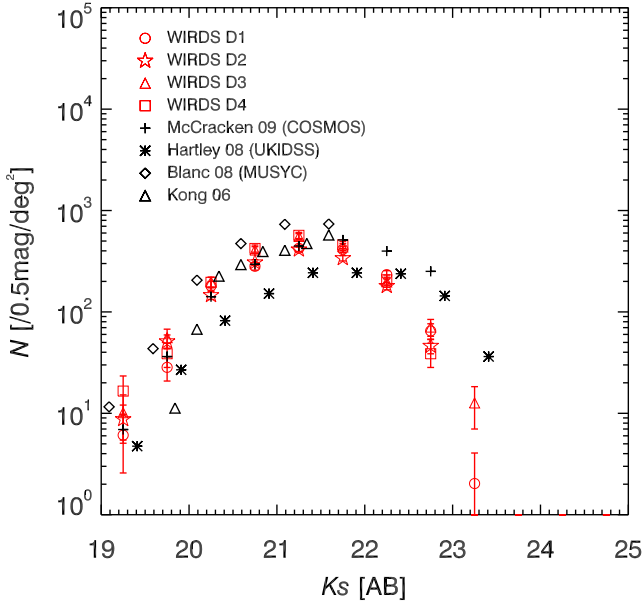}
   \caption{WIRDS number counts for p$BzK$ galaxies in the D1 (red circles), D2 (red stars), D3 (red triangles) and D4 fields (red squares). Literature counts are from the same sources and using the same symbols as Fig.~\ref{fig:numcntsBzK}.}
   \label{fig:numcntpBzK}
\end{figure}

The WIRDS number counts for the p$BzK$ samples are shown in Fig.~\ref{fig:numcntpBzK}. The literature counts are from the same sources as given in Fig.~\ref{fig:numcntsBzK}, with \citet{2006ApJ...638...72K} counts given by triangles, \citet{blanc08} counts denoted by diamonds, \citet{2008MNRAS.391.1301H} given by asterisks and the counts of \citet{2010ApJ...708..202M} given by crosses. Once again, the individual WIRDS fields are consistent with each other and are in good agreement with literature counts, except in the case of \citet{2008MNRAS.391.1301H}, which are somewhat lower than the other p$BzK$ counts. The WIRDS p$BzK$ number counts combined across all four WIRDS fields are given in Table~\ref{table:pbzkcounts}.

As noted by \citet{2010ApJ...708..202M}, a turn-over in the p$BzK$ number counts is seen at $K_{\rm s}\gtrsim21.5$, which indicate a lack of passive galaxies at $z\gtrsim2$. We note also that we see few passive galaxies at $z\gtrsim1.8$ identified by rest-frame colours in our photometric redshift sample (see Fig.~\ref{fig:bzk_nz}). This is partially due to the limiting magnitude of our optical data. For a typical passive galaxy, we expect a $z-K_{\rm s}\sim2.5$ and $B-z\sim4$, which for $K_{\rm s}\approx21.5$ would suggest optical magnitudes of $g\sim28$ and $z\sim24$. This is clearly a challenge for our $g$-band data, which have mean 50\% completion limits of $g=26.2$ / $g=26.7$ for extended/point-like objects in each of the four fields.

\begin{table}
\caption{WIRDS p$BzK$ number counts in the four WIRDS field}             
\label{table:pbzkcounts}      
\centering          
\begin{tabular}{l r r}    
\hline\hline       
$K_{\rm s}$ & $N_{\rm gal}$&$ N_{\rm gal}$  \\
(AB) & &  (/0.5mag/deg$^{2})$ \\
\hline                    
19.25 &      20 &       9.8 \\
19.75 &      91 &      44.8 \\
20.25 &     355 &     174.7 \\
20.75 &     699 &     344.0 \\
21.25 &     964 &     474.4 \\
21.75 &     814 &     400.6 \\
22.25 &     412 &     202.8 \\
22.75 &     111 &      54.6 \\
\hline                  
\end{tabular}
\end{table}

\section{Galaxy mass functions}
\label{sec:massfunc}

We now present an analysis of the galaxy mass function based on the WIRDS photometric $gri$ $\chi^2$-selected catalogues. The stellar mass function traces the build-up of the stellar mass content of galaxies, which provides a key observational constraint in modelling galaxy evolution and the physics that regulates it. With the deep NIR data and 2.4 deg$^2$ coverage over four fields, WIRDS provides an excellent dataset for such analysis, allowing the measurement of the mass function over a broad range of stellar mass, whilst the survey strategy significantly reduces the impact of cosmic variance on the results. 

\subsection{Estimating the WIRDS mass limits}

Before calculating the stellar mass functions of our galaxy populations, we first evaluate the mass limits of the galaxy population in the four WIRDS fields as a function of redshift. This is done following the method of \citet{2010ApJ...709..644I}, in which mass limits are calculated as the lowest mass at which less than $30\%$ of galaxies are fainter than a chosen magnitude limit. We estimate this mass limit as a function of galaxy redshift and type (i.e. passive, star-forming and both combined) using a magnitude limit of $i=25.5$.

\begin{figure}
\centering
\includegraphics[width=90.mm]{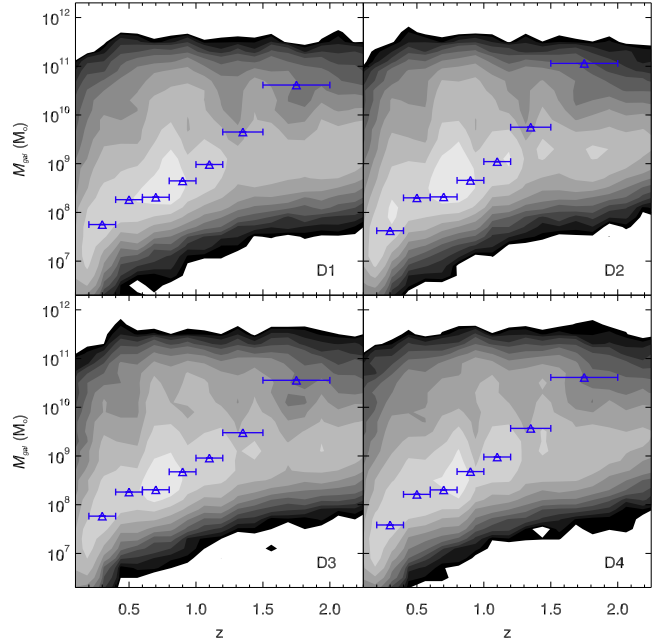}
   \caption{Estimated mass limits for the photometric catalogues in each of the four WIRDS fields. The greyscale contours show the distribution of the $i<25.5$ galaxy population, normalised by area. Blue triangles with horizontal error bars show the estimated mass limits for consecutive bins in redshift corresponding to a magnitude cut of $i=25.5$ for all galaxies.}
   \label{fig:mass_lim}
\end{figure}

\begin{table}
\caption{Mass limits based on a maximum 30\% of objects at magnitudes of $i>25$.5}             
\label{tab:masslimits}
\centering          
\begin{tabular}{cc}     
\hline\hline       
Redshift   & Mass Limit \\
           & ($\mbox{log}_{10}(M/M_{\odot})$) \\
\hline
$0.2<z\leq0.4$ & 7.75 \\
$0.4<z\leq0.6$ & 8.26 \\
$0.6<z\leq0.8$ & 8.31 \\
$0.8<z\leq1.0$ & 8.67 \\
$1.0<z\leq1.2$ & 9.98\\
$1.2<z\leq1.5$ & 9.65\\
$1.5<z\leq2.0$ & 10.61\\
\hline
\hline                  
\end{tabular}
\end{table}

\begin{figure*}
\centering
\includegraphics[width=160.mm]{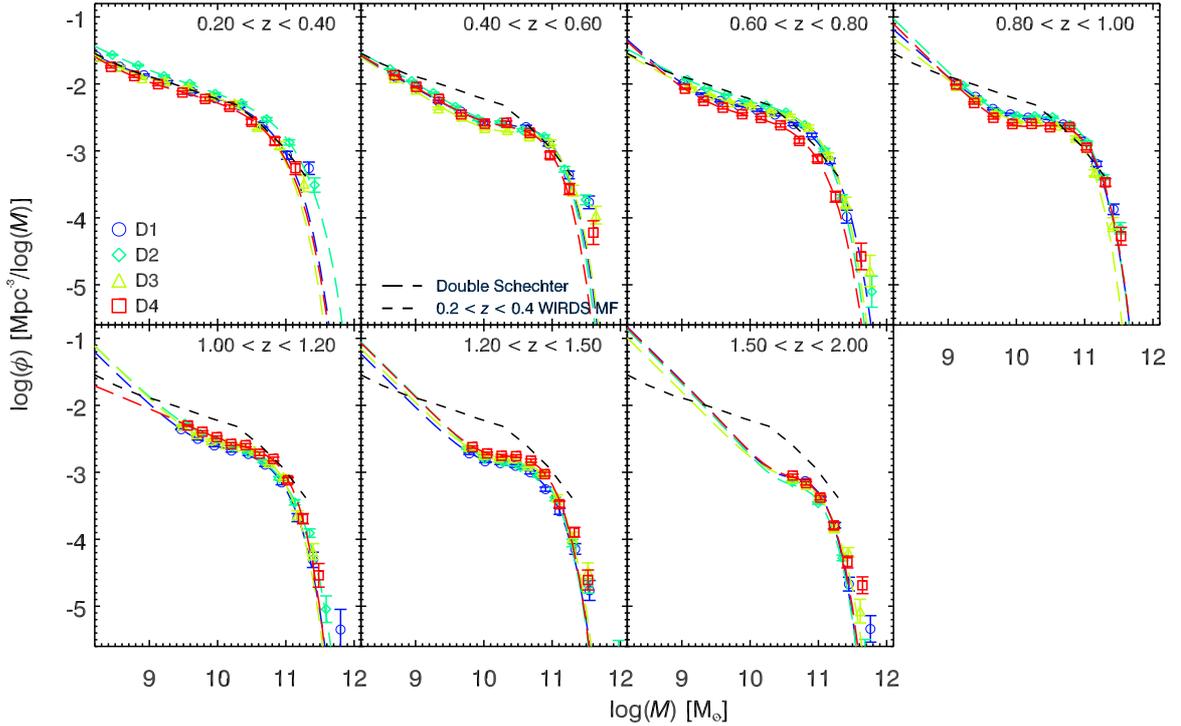}
   \caption{The mass function is shown as a function of redshift in each of the four WIRDS fields (blue circles, green diamonds, yellow triangles and red squares for the D1, D2, D3 and D4 fields respectively). Measurements in each field were performed using different mass bins in order to be presented clearly in the plotting (i.e. there is no artificial offset between the plotted mass functions). Errors on the points are a combination of the Poissonian errors with the uncertainties from the photometric redshift estimates. The double Schechter fits to the data are shown by the long-dashed curves, whilst the short-dashed line that shows the $0.2<z<0.4$ WIRDS mass function replicated in each panel for comparison.}
   \label{fig:mfunc}
\end{figure*}

Fig.~\ref{fig:mass_lim} shows the resulting mass limits as a function of redshift for each field. The grey-scale contours show the galaxy population distribution at $i<25.5$ normalised by area. Mass limits for the galaxy population are shown by the blue triangles. The extent of each redshift bin is given by the horizontal error bars. Estimated limits across the four fields are broadly consistent based on the imposed $i<25.5$ limit. In each of our five redshift bins we have taken the median mass limit of the four fields. These are given in Table~\ref{tab:masslimits}.

\subsection{The stellar mass function from $z=0.2$ to $z=2$}

We calculate the mass functions using the \texttt{ALF} (Algorithm for Luminosity Function; \citealt{2005A&A...439..863I}) tool with a step-wise maximum likelihood \citep[SWML; ][]{1988MNRAS.232..431E} estimator. We note that although we only present the SWML results here, the mass functions have  also been calculated using the non-parametric $1/V_{\rm max}$ \citep{1968ApJ...151..393S}, C$^+$ \citep{1997A&A...326..477Z} and STY \citep{1979ApJ...232..352S} methods and all are in agreement over the considered mass range.

Fig.~\ref{fig:mfunc} shows the total galaxy mass functions in redshift slices from $z=0.2$ to $z=2$. Blue circles show the results from the CFHTLS/WIRDS D1 field, green diamonds the D2 (COSMOS) field, yellow triangles the D3 field and red squares the D4 field. Each of the four CFHTLS/WIRDS fields are in good  agreement with the results of \citet{2010ApJ...709..644I} who derived mass functions for the full COSMOS field (using the full COSMOS filter set) and with the results of \citet{2006A&A...459..745F} from the GOODS-MUSIC sample. The WIRDS stellar mass functions shown here are presently one of the most robust measurements of the mass function at $z\gtrsim1$. Taking the mass limits estimated above, the results cover a broad range of stellar masses, whilst the incorporation of four distinct fields allows us to make a robust estimate of field-to-field scatter.

We note that, as observed by a number of previous authors \citep[e.g.][]{2008MNRAS.388..945B,2010ApJ...709..644I,2010A&A...523A..13P}, when the galaxy stellar mass function is measured over a broad mass range such as we have, it is not well fit by a Schechter function. In order to fit the observed mass functions, we therefore implement the five parameter double Schechter function fit implemented by \citet{2008MNRAS.388..945B}. This takes the form:

\begin{equation}
\phi(M)\mbox{d}M=e^{-M/M^\star}\left[\phi_1*\left(\frac{M}{M^\star}\right)^{\alpha_1}+\phi_2*\left(\frac{M}{M^\star}\right)^{\alpha_2}\right]\frac{\mbox{d}M}{M^\star}
\end{equation}

\noindent where $\phi(M)\mbox{d}M$ is the number density of galaxies with mass between $M$ and $M+\mbox{d}M$ and $M^\star$, $\alpha_1$, $\alpha_2$, $\phi_1^*$ and $\phi_2^*$ are the free parameters in the fit. We note that for the slopes, we maintain $\alpha_1>\alpha_2$ and that $\alpha_2\geq-2$, ensuring that the second term dominates at low-masses and that the luminosity density does not become divergent. The resultant double Schechter function fits are plotted (long-dashed curves) in Fig.~\ref{fig:mfunc} for each redshift/field combination. The double Schechter function fits consistently provide reliable fits to the mass function.

\begin{figure}
\centering
\includegraphics[width=90.mm]{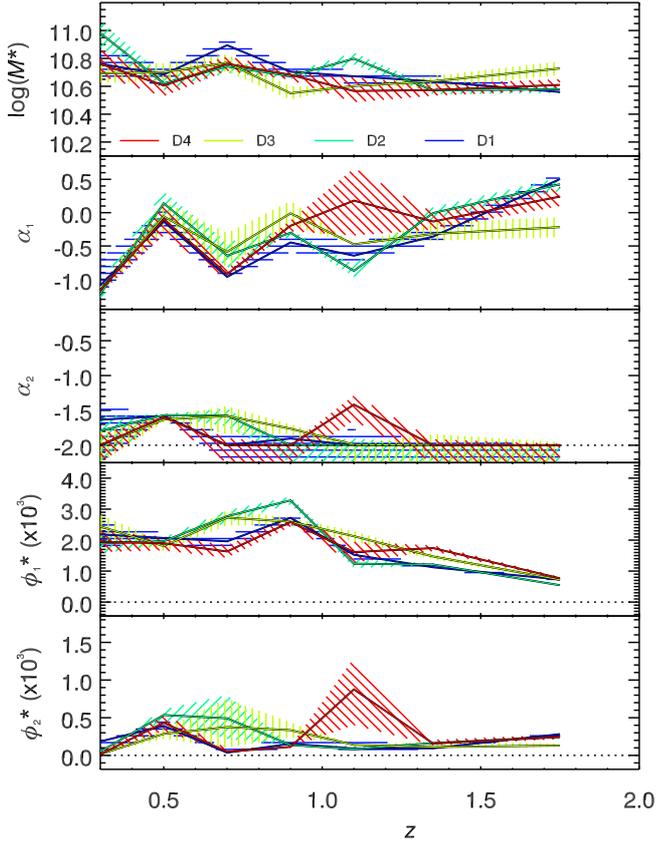}
   \caption{Evolution of the five parameters in the double Schechter function fits ($M^\star$, $\alpha_1$, $\alpha_2$, $\phi_1^\star$ and $\phi_2^\star$) as a function of redshift. The hatched regions show the $1\sigma$ uncertainty on the fit parameters.}
   \label{fig:mfparevo}
\end{figure}

In Fig.~\ref{fig:mfparevo} we show the double Schechter function fit parameters as a function of redshift for our sample. The blue, green, yellow and red lines in each panel show the parameters for the fields D1, D2, D3 and D4 respectively, whilst the hatched regions show the uncertainties on the parameters in each case. The parameter values are also given in Table~\ref{tab:schechter_params}.

\begin{figure}
\centering
\includegraphics[width=90.mm]{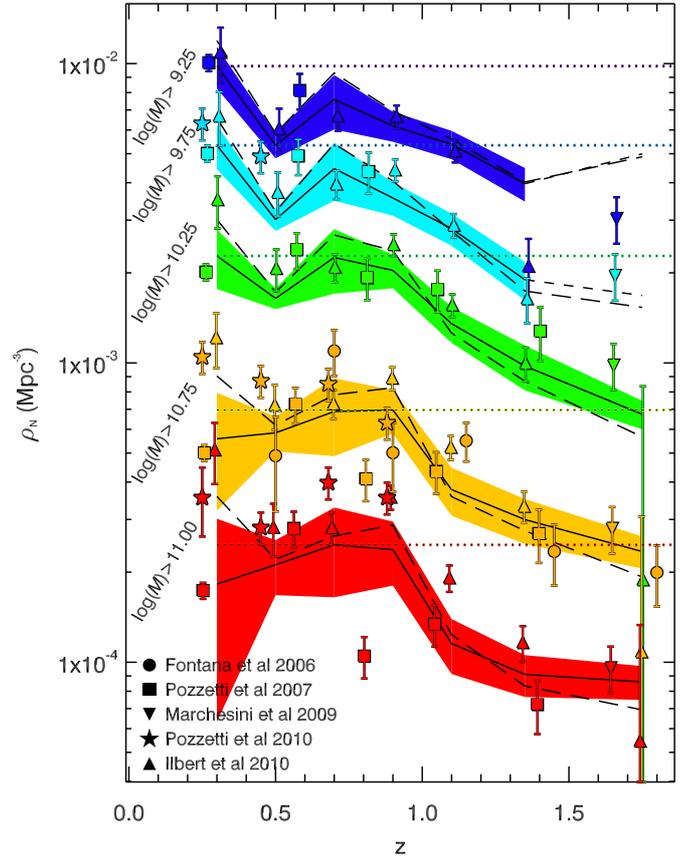}
   \caption{Evolution of the galaxy number density as a function of redshift for the mass limits $\mbox{log}(M/M_\odot)>9.0$, $\mbox{log}(M/M_\odot)>9.5$, $\mbox{log}(M/M_\odot)>10.0$, $\mbox{log}(M/M_\odot)>10.5$ and $\mbox{log}(M/M_\odot)>11.0$. For each mass limit, the solid line gives the mean number density across the four fields, whilst the shaded region gives the scatter between the fields. Dashed lines give the measurements that are incomplete based on our mass-completeness estimates. Below $z\sim1$, the numbers of massive galaxies remain relatively constant, whilst the numbers of low-mass galaxies increase with decreasing redshift. Above $z\sim1$, the build-up of massive galaxies is seen with a significant increase in galaxies of masses $\mbox{log}(M/M_\odot)>10.5$ and $\mbox{log}(M/M_\odot)>11.0$ evident from $z\sim2$ to $z\sim1$.}
   \label{fig:numdens}
\end{figure}

Using the Schechter function fits, we calculate the galaxy number densities for given minimum mass cuts as a function of redshift. The results are shown in Fig.~\ref{fig:numdens}, where we show the number densities for mass cuts of $M>10^{9.25}M_\odot$ (blue), $M>10^{9.75}M_\odot$ (cyan), $M>10^{10.25}M_\odot$ (green), $M>10^{10.75}M_\odot$ (yellow) and $M>10^{11}M_\odot$ (red). In each case, the solid lines give the mean number density across the four fields, with the shaded regions giving the scatter ($1\sigma$) between the four fields. Short-dashed lines denote results where $M_{min}$ is less than the mass completeness limit for a given redshift. We compare our results to a number of other surveys denoted by the points with error bars. These are calculated in the same manner as the WIRDS data using the published Schechter fits for \citet[][squares]{2007A&A...474..443P}, \citet[][upside-down triangles]{2009ApJ...701.1765M} and \citet[][triangles]{2010ApJ...709..644I} and from the published points in the case of \citet[][circles]{2006A&A...459..745F} and \citet[][stars]{2010A&A...523A..13P}. The results of \citet{2010ApJ...709..644I} and \citet[][stars]{2010A&A...523A..13P} and those from the WIRDS D2 field (long dashed lines) are all obtained from the same field (COSMOS), although with slightly differing datasets. It is promising to see that these are all consistent with each other as shown in Fig.~\ref{fig:numdens}. Additionally, an important result from this analysis is that we see the field-to-field variation between the WIRDS results in the four fields, which is highlighted by the high values for the number density in the COSMOS field at redshifts of $z\sim0.3$ and $z\sim0.6-1$. These high densities are seen in all three D2/COSMOS measurements and relate to known large scale over-densities in the field \citep{2007ApJS..172..182F,2007ApJS..172..314M,2009A&A...505..463M}. By combining the four WIRDS fields, we see a more accurate picture of the stellar mass content traced over cosmic time.

The WIRDS data are consistent with mass assembly downsizing in which the more massive galaxy populations (i.e. $M\gtrsim10^{10.75}M_\odot$) are mostly in place by $z\sim1$, whilst the numbers of lower mass objects are still increasing with lower redshifts. For instance, increases of 0.19 dex, 0.19 dex and 0.08 dex are seen for the mass limits of $10^{9.25}M_\odot$, $10^{9.75}M_\odot$ and $10^{10.25}M_\odot$ respectively from $z=0.8$ to $z=0.3$. No increase is evident in the number densities between $z=0.8$ to $z=0.3$ in the higher mass samples. Comparing to data from the literature, we find a good consistency between the datasets, in particular for the lower mass cuts. At higher masses, the range of data consistently shows the number of massive galaxies reaching a maximum at $z\sim0.7-1.2$. Although the size of our fields means we are unable to probe much higher in mass than $\sim10^{11}M_\odot$ in this way, we note that other authors have reported this trend continuing with higher mass galaxy populations having reached their present day numbers at ever higher redshift \citep[e.g.][]{2011MNRAS.413.2845M}.

\begin{figure*}
\centering
\includegraphics[width=180.mm]{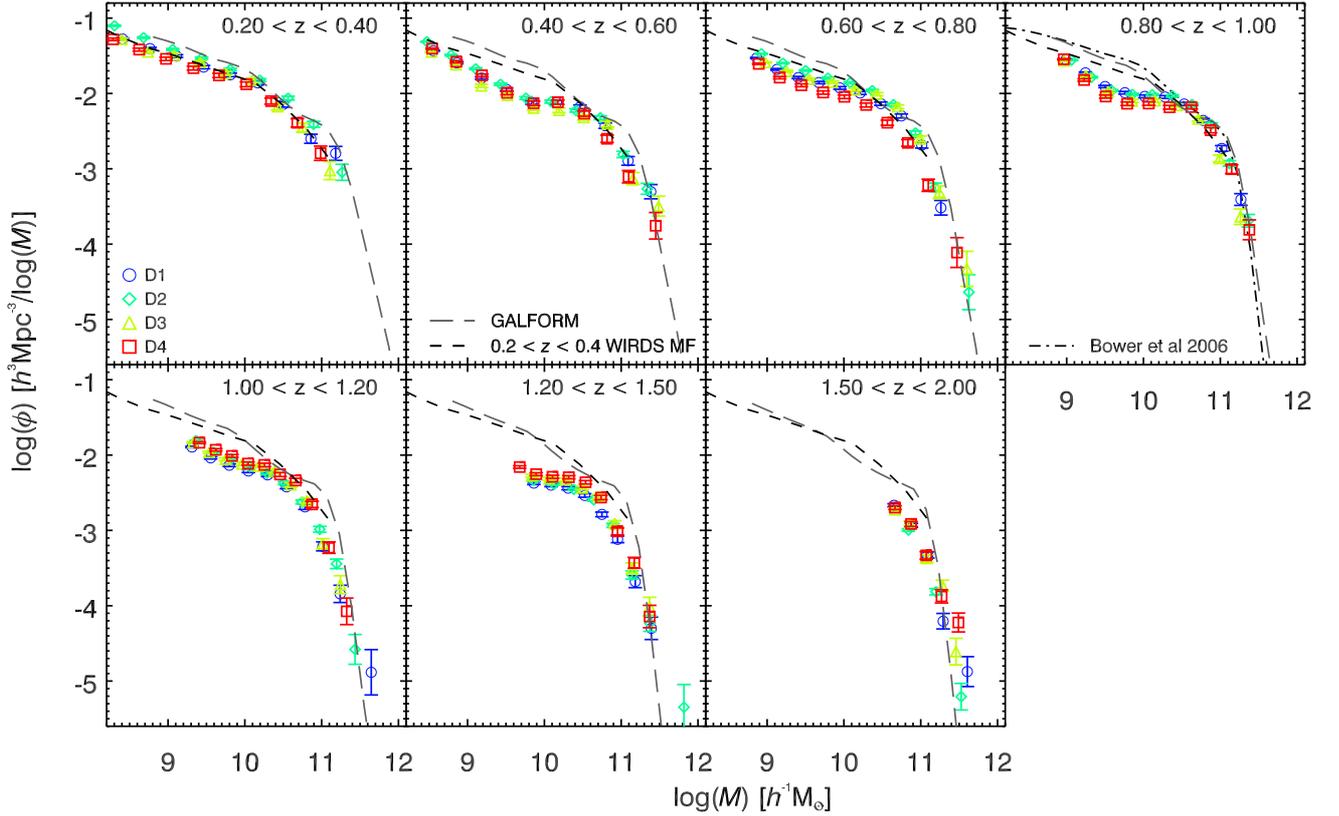}
   \caption{The observed mass functions are reproduced as in Fig.~\ref{fig:mfunc} (blue circles, green diamonds, yellow triangles and red squares for the D1, D2, D3 and D4 fields respectively), but now with the GALFORM results as comparison. The solid line in each redshift range shows the equivalent result in each redshift range based on the GALFORM simulation results, whilst the dashed line again shows the $0.2<z<0.4$ WIRDS mass function repeated from the first panel. The dash-dot line in the $0.8<z<1.0$ shows the original $z=1$ mass function directly from \citet{2006MNRAS.370..645B}.}
   \label{fig:mfunc_gf}
\end{figure*}

This lack of evolution in the massive galaxy population is consistent with studies of brightest cluster galaxies (BCGs) in which little change is observed in BCG stellar masses to redshifts of $z\sim1-1.5$ \citep{2010ApJ...718...23S}. These observations corroborate the picture of galaxy evolution whereby a mass-dependent shut-down of star formation is experienced by galaxies \citep[e.g.][]{2008ApJS..175..356H,2008ApJS..175..390H,2009ApJ...696..891H}. The cause of this quenching of star-formation is uncertain, but likely candidates are that the gas fuelling star-formation is exhausted by heavy star-formation; that the gas is heated to temperatures beyond which it may collapse to form stars due to the high virial temperature of the host halo; or that the gas is expelled by via feedback from the central AGN.

\subsection{Comparison to simulations}

We compare our measurements with predictions from the \citet{2006MNRAS.370..645B} implementation of the ``GALFORM'' semi-analytic simulations which are shown as the solid curves in Fig.~\ref{fig:mfunc_gf}. The results of GALFORM have been compared to previous observational data by \citeauthor{2006MNRAS.370..645B} who show good fits to the observed mass functions of \citet{2004Natur.430..181G}, \citet{2004A&A...424...23F} and \citet{2005ApJ...619L.131D}. However we note that the WIRDS results provide a much stronger constraint on the mass function over the $1<z<2$ redshift range than these previous results. Some of the key features of the \citeauthor{2006MNRAS.370..645B} GALFORM implementation are (i) a time scale for quiescent star formation that varies with the dynamical time of the disk and which therefore changes significantly with redshift, (ii) bursts of star formation occur due to both galaxy mergers and when disks become dynamically unstable, and (iii) the inclusion of both supernova and active galactic nuclei (AGN) feedback. This feedback is implemented in such a way that active galactic nucleii (AGNs) are able to heat the cooling flows in massive haloes, quenching star formation. \citeauthor{2006MNRAS.370..645B} adopt the cosmological parameters of the Millennium Simulation \citep{2005Natur.435..629S}, consistent with cosmic microwave background radiation and large scale galaxy clustering \citep[e.g.][]{sanchez09}: $\Omega_{0}=0.25$, $\Lambda_{0} = 0.75$, $\Omega_{b}=0.045$, $\sigma_{8}=0.9$ and $h=0.73$. The \citeauthor{2006MNRAS.370..645B} model parameters were fixed with reference to a subset of the available observations of galaxies, mostly at low redshift. For further details we refer the reader to \citeauthor{2006MNRAS.370..645B}.

We note that GALFORM uses a Kennicutt IMF, whilst the photometric masses derived from the WIRDS data assume a Chabrier IMF. Based on \citet{2010ApJ...709..644I} and \citet{2011MNRAS.414..304G}, we multiply the GALFORM masses by a factor of 1.32 to match the Chabrier based stellar masses of the WIRDS photometric catalogues.  The \citeauthor{2006MNRAS.370..645B} GALFORM results have been shown to successfully reproduce the stellar mass function up to $z=4.5$ and the number counts of red galaxies at $z<2$ \citep{2008MNRAS.386.2145A,2009MNRAS.398..497G}.

The GALFORM mass functions as a function of redshift are plotted alongside the WIRDS results in Fig.~\ref{fig:mfunc_gf} (solid curves). The dash-dot line in the $0.8<z<1.0$ panel shows the original \citeauthor{2006MNRAS.370..645B} $z=1$ mass function\footnote{We note that the shape of the mass function predicted with the \citeauthor{2006MNRAS.370..645B} model presented here differs slightly from that in the original paper. After the publication of \citeauthor{2006MNRAS.370..645B} a bug was corrected, forcing slight changes in a number of parameters, affecting the shape of the mass function.}. Here we see that the GALFORM model shows some success in reproducing the CFHTLS WIRDS mass functions up to $z\approx2$, showing good agreement at intermediate masses with both the overall numbers and the position of the break in the mass function. We note however some deviations of the model from the observations, with the counts being over-predicted at $\mbox{log}(M/M_\odot)\lesssim10$, whilst the model begins to generally under-predict the numbers of galaxies at masses of $\mbox{log}(M/M_\odot)\sim11$, especially at redshifts of $z\gtrsim1$. The over-prediction of galaxies with masses of $\mbox{log}(M/M_\odot)\lesssim10$ may be attributed to the form and strength of the supernova-feedback prescriptions in the simulation, as discussed in \citet{2012MNRAS.422.2816B}.

We also note that the models show an evolution with redshift in the slope of the mass function above $M^\star$, suggesting the rate of increase in numbers of galaxies with decreasing redshift increases as a function of stellar mass above $M^\star$. It is not clear that this is seen in the data where the slope at $M\gtrsim M^\star$ appears relatively consistent across the redshift slices where it is clearly defined. We also note that the GALFORM results are not convolved with the observational errors and that this may reduce the tension in terms of the evolution or lack thereof of the high mass slopes.

\section{Conclusions}
\label{sec:conclusions}

\begin{table*}
\caption{Double Schechter function fit parameters (and associated 1$\sigma$ uncertainties) for each field as a function of redshift.}
\centering
\label{tab:schechter_params}
\begin{tabular}{llccccccc}
\hline\hline
                   Parameter &  Field &              $z = 0.30$ &              $z = 0.50$ &              $z = 0.70$ &              $z = 0.90$ &              $z = 1.10$ &              $z = 1.35$ &              $z = 1.75$ \\
\hline
\hline
                  $\alpha_1$ &     D1 & $-1.09^{+0.41}_{-0.15}$ & $-0.13^{+0.20}_{-0.20}$ & $-0.96^{+0.09}_{-0.05}$ & $-0.45^{+0.15}_{-0.14}$ & $-0.65^{+0.21}_{-0.10}$ & $-0.36^{+0.11}_{-0.11}$ & $ 0.50^{+0.11}_{-0.11}$ \\
                             &     D2 & $-1.22^{+0.17}_{-0.07}$ & $ 0.14^{+0.18}_{-0.18}$ & $-0.64^{+0.20}_{-0.18}$ & $-0.30^{+0.05}_{-0.05}$ & $-0.87^{+0.09}_{-0.09}$ & $-0.00^{+0.08}_{-0.08}$ & $ 0.43^{+0.10}_{-0.10}$ \\
                             &     D3 & $-1.16^{+0.07}_{-0.04}$ & $-0.06^{+0.23}_{-0.22}$ & $-0.59^{+0.28}_{-0.23}$ & $-0.01^{+0.19}_{-0.18}$ & $-0.47^{+0.00}_{+0.00}$ & $-0.32^{+0.18}_{-0.10}$ & $-0.22^{+0.15}_{-0.15}$ \\
                             &     D4 & $-1.16^{+0.07}_{-0.04}$ & $-0.10^{+0.25}_{-0.24}$ & $-0.92^{+0.08}_{-0.07}$ & $-0.20^{+0.08}_{-0.08}$ & $ 0.18^{+0.49}_{-0.52}$ & $-0.13^{+0.13}_{-0.12}$ & $ 0.24^{+0.14}_{-0.14}$ \\
\hline
                  $\alpha_2$ &     D1 & $-1.63^{+0.20}_{+0.20}$ & $-1.58^{+0.03}_{-0.04}$ & $-2.00^{+0.16}_{+0.16}$ & $-1.90^{+0.09}_{+0.09}$ & $-2.00^{+0.23}_{+0.23}$ & $-2.00^{+0.06}_{+0.06}$ & $-2.00^{+0.01}_{+0.01}$ \\
                             &     D2 & $-1.79^{+0.21}_{+0.21}$ & $-1.57^{+0.02}_{-0.02}$ & $-1.57^{+0.10}_{-0.15}$ & $-2.00^{+0.02}_{+0.02}$ & $-2.00^{+0.10}_{+0.10}$ & $-2.00^{+0.01}_{+0.01}$ & $-2.00^{+0.01}_{+0.01}$ \\
                             &     D3 & $-2.00^{+0.22}_{+0.22}$ & $-1.63^{+0.04}_{-0.04}$ & $-1.58^{+0.14}_{-0.26}$ & $-1.76^{+0.06}_{-0.07}$ & $-1.99^{+0.00}_{+0.00}$ & $-2.00^{+0.16}_{+0.16}$ & $-2.00^{+0.05}_{+0.05}$ \\
                             &     D4 & $-2.00^{+0.15}_{+0.15}$ & $-1.59^{+0.04}_{-0.04}$ & $-2.00^{+0.10}_{+0.10}$ & $-2.00^{+0.02}_{+0.02}$ & $-1.42^{+0.13}_{-0.23}$ & $-2.00^{+0.08}_{+0.08}$ & $-2.00^{+0.02}_{+0.02}$ \\
\hline
                  log$(M^*)$ &     D1 & $10.76^{+0.11}_{-0.13}$ & $10.68^{+0.07}_{-0.06}$ & $10.89^{+0.05}_{-0.05}$ & $10.70^{+0.05}_{-0.05}$ & $10.67^{+0.04}_{-0.05}$ & $10.63^{+0.04}_{-0.04}$ & $10.56^{+0.03}_{-0.03}$ \\
                 $(M_\odot)$ &     D2 & $10.98^{+0.09}_{-0.10}$ & $10.62^{+0.05}_{-0.05}$ & $10.74^{+0.04}_{-0.04}$ & $10.67^{+0.02}_{-0.02}$ & $10.80^{+0.04}_{-0.04}$ & $10.57^{+0.03}_{-0.03}$ & $10.58^{+0.03}_{-0.03}$ \\
                             &     D3 & $10.69^{+0.07}_{-0.07}$ & $10.70^{+0.06}_{-0.06}$ & $10.77^{+0.06}_{-0.06}$ & $10.55^{+0.05}_{-0.05}$ & $10.60^{+0.00}_{+0.00}$ & $10.64^{+0.04}_{-0.04}$ & $10.73^{+0.05}_{-0.05}$ \\
                             &     D4 & $10.76^{+0.10}_{-0.08}$ & $10.61^{+0.08}_{-0.08}$ & $10.76^{+0.05}_{-0.05}$ & $10.68^{+0.04}_{-0.03}$ & $10.57^{+0.10}_{-0.08}$ & $10.57^{+0.04}_{-0.04}$ & $10.61^{+0.04}_{-0.04}$ \\
\hline
                  $\phi_1^*$ &     D1 & $ 2.18^{+0.38}_{-0.37}$ & $ 2.07^{+0.17}_{-0.20}$ & $ 1.96^{+0.21}_{-0.20}$ & $ 2.72^{+0.19}_{-0.21}$ & $ 1.53^{+0.14}_{-0.15}$ & $ 1.12^{+0.08}_{-0.09}$ & $ 0.73^{+0.03}_{-0.03}$ \\
$(\times10^{-3}$ Mpc$^{-3})$ &     D2 & $ 1.90^{+0.33}_{-0.32}$ & $ 1.96^{+0.11}_{-0.12}$ & $ 2.77^{+0.19}_{-0.20}$ & $ 3.28^{+0.12}_{-0.12}$ & $ 1.24^{+0.13}_{-0.13}$ & $ 1.21^{+0.04}_{-0.04}$ & $ 0.54^{+0.02}_{-0.02}$ \\
                             &     D3 & $ 2.44^{+0.38}_{-0.35}$ & $ 1.81^{+0.14}_{-0.16}$ & $ 2.72^{+0.23}_{-0.25}$ & $ 2.60^{+0.13}_{-0.13}$ & $ 2.12^{+0.16}_{-0.18}$ & $ 1.47^{+0.09}_{-0.09}$ & $ 0.72^{+0.05}_{-0.06}$ \\
                             &     D4 & $ 1.94^{+0.36}_{-0.33}$ & $ 1.90^{+0.21}_{-0.23}$ & $ 1.63^{+0.21}_{-0.20}$ & $ 2.58^{+0.14}_{-0.15}$ & $ 1.61^{+0.33}_{-0.34}$ & $ 1.74^{+0.08}_{-0.10}$ & $ 0.77^{+0.04}_{-0.04}$ \\
\hline
                  $\phi_2^*$ &     D1 & $ 0.19^{+0.00}_{-0.18}$ & $ 0.39^{+0.09}_{-0.09}$ & $ 0.04^{+0.05}_{-0.01}$ & $ 0.15^{+0.08}_{-0.06}$ & $ 0.09^{+0.12}_{-0.02}$ & $ 0.09^{+0.02}_{-0.01}$ & $ 0.28^{+0.02}_{-0.02}$ \\
$(\times10^{-3}$ Mpc$^{-3})$ &     D2 & $ 0.05^{+0.00}_{+0.00}$ & $ 0.54^{+0.07}_{-0.07}$ & $ 0.49^{+0.31}_{-0.27}$ & $ 0.14^{+0.01}_{-0.01}$ & $ 0.08^{+0.04}_{-0.02}$ & $ 0.16^{+0.01}_{-0.01}$ & $ 0.25^{+0.02}_{-0.02}$ \\
                             &     D3 & $ 0.02^{+0.06}_{-0.00}$ & $ 0.29^{+0.07}_{-0.07}$ & $ 0.38^{+0.38}_{-0.28}$ & $ 0.34^{+0.10}_{-0.09}$ & $ 0.14^{+0.00}_{+0.00}$ & $ 0.11^{+0.08}_{-0.01}$ & $ 0.13^{+0.03}_{-0.02}$ \\
                             &     D4 & $ 0.02^{+0.04}_{-0.00}$ & $ 0.43^{+0.11}_{-0.11}$ & $ 0.05^{+0.03}_{-0.01}$ & $ 0.11^{+0.01}_{-0.01}$ & $ 0.88^{+0.42}_{-0.47}$ & $ 0.16^{+0.05}_{-0.02}$ & $ 0.24^{+0.03}_{-0.03}$ \\
\hline
\hline
\end{tabular}
\end{table*}

In this paper we have presented a new deep, $JHK_{\rm s}$ near-infrared survey using WIRCam which partially covers the four $1\deg2$ CFHTLS deep fields. The addition of high-quality near-infrared data reaching $K\sim24$ (AB) is an essential tool for addressing a number of crucial outstanding questions in galaxy formation and evolution including, but not limited to, a precise measurement of luminosity and stellar mass functions at $z>1$ \citep[e.g.][]{2004A&A...424...23F,2006A&A...459..745F,2007MNRAS.380..585C,2010ApJ...709..644I}; investigating the evolutionary history of the most massive galaxies beyond $z\sim3$ \citep[e.g. ][]{2006A&A...459..745F} and probing the changing relationship between baryons and dark matter beyond redshift one \citep[e.g.][]{2010MNRAS.406..147F,2010MNRAS.407.1212H,2011ApJ...728...46W}. Furthermore, deep, wide, near-infrared data facilitates the identification of high-redshift groups and clusters \citep[e.g.][]{2007ApJS..172..182F,2010A&A...523A..66B,2010MNRAS.403.2063F,2010ApJ...725..615H,2011A&A...526A.133G,2011NJPh...13l5014F} and the first populations of galaxies in the early Universe \citep[e.g.][]{2006Natur.440.1145H,2010ApJ...723..869O,2011ApJ...730...68C}.

The final images cover areas of $0.49\deg2$, $0.78\deg2$, $0.40\deg2$ and $0.36\deg2$ in the D1, D2/COSMOS, D3/AEGIS and D4/LBQS2212-17 fields respectively, giving a total effective area of $2.03\deg2$. Measured full-width-half maxima in all fields are $\approx0.8\arcsec$ or better (except for the D2/COSMOS WFCAM $J$-band image, which has a FWHM $\sim1.0\arcsec$. For the D1, D2 and D4 CFHTLS fields, the depths in all three bands are consistently better than mag$_{AB}\sim24.5$ ($50\%$ point-source completeness), whilst the D2/COSMOS field is slightly shallower reaching $J=23.4$, $H=24.1$ and $K_{\rm s}=24.0$. We have presented combined counts for our $2.03\deg2$ survey in all three bands. These are consistent with previous works, whilst our measurements are generally either made over a larger area or deeper than previous measurements.

WIRDS data have been combined with CFHTLS deep optical data in each of our four fields to create high quality 8-band photometric catalogues. From the full 8-band $ugrizJHK_{\rm s}$ data-set we have constructed two sets of combined catalogues, the first catalogue set made using $\chi^2$ $gri$ detection images and the second set constructed using $K_{\rm s}$ images as the detection source. All WIRDS images and catalogues are publicly available to download from the CADC. Photometric redshifts have been determined based on both of these catalogue sets using the Le PHARE software. In addition we have produced a catalogue of galaxy properties based on the SED fitting using Le PHARE using the $\chi^2$ $gri$ based catalogue, giving estimates of galaxy stellar masses.

We have presented an analysis of the success of the $BzK$ selection based on the CFHTLS/WIRDS data compared to full photometric SED analyses for redshift and galaxy type determination. Our analysis has shown that the p$BzK$ selection based on the CFHTLS/WIRDS filters successfully selects $\approx52\%$ of the passive galaxy population at $1.4<z<2$ when compared to the photometric SED-fitted catalogue. In addition, the selection selects a significant number of $1<z<1.4$ galaxies. The s$BzK$ selection successfully identifies $76\%$ of the $1.4<z<2$ star-forming galaxy population identified by our SED fitting, with a much stronger cut at $z=1.4$ than seen for the p$BzK$ selection.

We have presented estimated mass limits for our $i$-selected photometric catalogue and presented the total galaxy mass function as a function of redshift over the range $0.2 < z < 2$. These results provide one of the most robust measurements of the galaxy stellar mass function at $z\gtrsim1$ presently available. In particular, we have shown that the mass functions are consistent between each of the four individual fields covered in this survey, providing a strong constraint on the effect of cosmic variance on our measurements. The results are consistent with the current best determinations of the stellar mass function over our redshift range. We have shown that the mass functions are well fit by double Schechter function fits, whilst noting that single Schechter functions do not provide good fits across the broad mass ranges covered. The evolution of the Schechter function parameters have been presented as a function of redshift for each of the four fields, noting in particular that $M^\star$ shows a gradual decline from $z=0$ to $z=2$. From the fits to the mass functions, we have calculated the number densities of galaxies above a set of mass limits as a function of redshift. The results are consistent between the for fields and show the build-up of mass as a function of redshift since $z=2$. We find that since $z\sim0.8-1$, there has been little change in the numbers of massive galaxies ($M\gtrsim10^{10.75}$), whilst the formation of lower mass galaxies has been ongoing in the same epoch. This is consistent with the previous findings of for example \citet{2007A&A...474..443P,2010A&A...523A..13P,2010ApJ...709..644I}. From $z\sim2$ to $z\sim1$, the WIRDS data clearly show the significant increase in numbers of galaxies of all masses probed by the survey as the Universe undergoes the phase of peak star-forming activity \citep{1996APJ...460L...1L,1996MNRAS.283.1388M}.

Finally, we have compared our results with the predictions of the semi-analytical galaxy formation model GALFORM and find that the simulations provide a relatively successful fit to the observed mass functions at intermediate masses of $10\lesssim{\rm log}(M/M_\odot)\lesssim11$. However, as is common with semi-analytical predictions of the mass function, the GALFORM results under-predict the mass function at low masses (i.e. ${\rm log}(M/M_\odot)\lesssim10$), whilst the fit as a whole degrades beyond redshifts of $z\sim1.2$.

\begin{acknowledgements}
RMB acknowledges the funding of the French Agence Nationale de la Recherche (ANR) and the UK Science and Technology Facilities Council. This work is based in part on data products produced at TERAPIX at the Institut d'Astrophysique de Paris and the Canadian Astronomy Data Centre (CADC) as part of the Canada-France-Hawaii Telescope (CFHT) Legacy Survey, a collaborative project of the Canadian National Research Council (NRC) and Centre National de la Recherche Scientifique (CNRS). Much of the work is based on observations at the CFHT, which is funded by the NRC, the CNRS, and the University of Hawaii. This research has made use of the VizieR catalogue access tool provided by the CDS, Strasbourg, France. This research was supported by ANR grant ``ANR-07-BLAN-0228''. ED also acknowledges support from ``ANR-08-JCJC-0008''.  JPK acknowledges support from the CNRS.
\end{acknowledgements}

\bibliographystyle{aa}
\bibliography{rmb}

\end{document}